\documentclass{aa}
\usepackage{rotating}
\usepackage{graphicx}
\usepackage{txfonts}
\begin{document}

\title{A spectroscopic investigation of early-type stars in the young open cluster Westerlund\,2\thanks{Based on observations collected at the European Southern Observatory (Cerro Paranal, Chile)}}
\author{G.\ Rauw\inst{1}\fnmsep\thanks{Honorary Research Associate FRS-FNRS (Belgium)} \and H.\ Sana\inst{2} \and Y.\ Naz\'e\inst{1}\fnmsep\thanks{Research Associate FRS-FNRS (Belgium)}}
\offprints{G.\ Rauw}
\mail{rauw@astro.ulg.ac.be}
\institute{Groupe d'Astrophysique des Hautes Energies, Institut d'Astrophysique et de G\'eophysique, Universit\'e de Li\`ege, All\'ee du 6 Ao\^ut, B\^at B5c, 4000 Li\`ege, Belgium 
\and Sterrenkundig Instituut `Anton Pannekoek', Universiteit van Amsterdam, Science Park 904, 1098 XH Amsterdam, The Netherlands} 
%European Southern Observatory, Alonso de Cordova 3107, 19 Vitacura, Santiago, Chile \and Observatoire de Grenoble, 414, Rue de la Piscine, BP 53, 38041 Grenoble, France \and D\'epartement de Physique, Universit\'e de Montr\'eal, QC, H3C 3J7, and Observatoire du Mont M\'egantic, Canada}
\date{Received date / Accepted date}
\abstract{The distance of the very young open cluster \object{Westerlund\,2}, which contains the very massive binary system \object{WR\,20a} and is likely associated with a TeV source, has been the subject of much debate.}{We attempt a joint analysis of spectroscopic and photometric data of eclipsing binaries in the cluster to constrain its distance.}{A sample of 15 stars, including three eclipsing binaries (\object{MSP\,44}, \object{MSP\,96}, and \object{MSP\,223}) was monitored with the FLAMES multi-object spectrograph. The spectroscopic data are analysed together with existing $B\,V$ photometry.}{The analysis of the three eclipsing binaries clearly supports the larger values of the distance, around 8\,kpc, and rules out values of about 2.4 -- 2.8\,kpc that have been suggested in the literature. Furthermore, our spectroscopic monitoring reveals no clear signature of binarity with periods shorter than 50\,days in either the WN6ha star \object{WR\,20b}, the early O-type stars \object{MSP\,18}, \object{MSP\,171}, \object{MSP\,182}, \object{MSP\,183}, \object{MSP\,199}, and \object{MSP\,203}, or three previously unknown mid O-type stars. The only newly identified candidate binary system is \object{MSP\,167}. The absence of a binary signature is especially surprising for WR\,20b and MSP\,18, which were previously found to be bright X-ray sources.} {The distance of Westerlund\,2 is confirmed to be around 8\,kpc as previously suggested based on the spectrophotometry of its population of O-type stars and the analysis of the light curve of WR\,20a. Our results suggest that short-period binary systems are not likely to be common, at least not among the population of O-type stars in the cluster.}
\authorrunning{Rauw, Sana \& Naz\'e}
\titlerunning{Early-type stars in Westerlund\,2}
\maketitle
\section{Introduction}
The open cluster Westerlund\,2 lies in a blowout region of the giant H\,{\sc ii} complex RCW\,49. Interest in the stellar population of Westerlund\,2 was triggered by two independent observational studies. On the one hand, RCW\,49 was observed with the Infrared Array Camera aboard {\it Spitzer} in the framework of the GLIMPSE legacy survey (Churchwell et al.\ \cite{Churchwell04}). These observations revealed strong evidence of ongoing star formation activity and underlined the need for a detailed study of the properties of the early-type stars in the cluster core and their impact on the surrounding nebula. On the other hand, WR\,20a - one of the two Wolf-Rayet stars in RCW\,49 - was found to be a very massive eclipsing binary consisting of two WN6ha stars with individual masses of about $80\,M_{\odot}$ (Rauw et al.\ \cite{Rauw04,Rauw05}; Bonanos et al.\ \cite{bonanos}). Since previous spectroscopic studies revealed only one O6: and six O7: stars, in the cluster core (Moffat, Shara \& Potter \cite{MSP}), the existence of a pair of stars that massive was somewhat unexpected. 

The nature of Westerlund\,2 thus remains to be established: is it a super-cluster maybe harbouring dozens of very massive early O-type stars, or is it "just" a medium-size agglomerate that produced the very massive WN6ha + WN6ha binary system WR\,20a more or less by chance? \footnote{Indeed, the existence of a relation between the mass of an open cluster and that of its most massive member remains disputed (see Weidner, Kroupa \& Bonnell \cite{Weidner} and references therein).} This question called for a re-investigation of the population of early-type stars in this cluster. A first step in this direction was performed by Rauw et al.\ (\cite{GR}, hereafter Paper I), who presented a photometric monitoring of the cluster as well as a spectroscopic snapshot study of its brightest members. This study allowed us to discover three previously unknown eclipsing binaries among the likely cluster members and to derive spectral types between O3 and O6.5 (i.e.\ significantly earlier than previously thought) for the twelve brightest O-stars in the cluster. Moreover, we inferred a spectrophotometric distance of $8.0 \pm 1.4$\,kpc. While this distance measurement is in excellent agreement with that of WR\,20a (Rauw et al.\ \cite{Rauw05}), it is nevertheless problematic in several ways. Indeed, despite the much earlier spectral types inferred for the most massive cluster members, the current census of early-type stars in Westerlund\,2 accounts for only 20\% of the ionizing photons required to explain the observed radio emission of the RCW\,49 complex. Moreover, a near-IR photometric study of the cluster by Ascenso et al.\ (\cite{Ascenso}) resulted in a distance estimate of 2.8\,kpc, inferred from a comparison between their $JHK_s$ data with pre-main sequence evolutionary tracks. The origin of these discrepancies remains unclear. More recently, yet another distance determination was obtained. From the mean velocity and velocity spread of molecular gas in the molecular cloud associated with RCW\,49, Furukawa et al.\ (\cite{Furukawa}) inferred a kinematic distance of $5.4^{+1.1}_{-1.4}$\,kpc. These authors further suggested that the collision between two molecular clouds might have triggered the formation of the Westerlund\,2 cluster. 

\begin{table*}[t!]
\caption{Overview of the targets studied in this paper.\label{journal}}
\begin{center}
\begin{tabular}{c c c c c c}
\hline
Object  & Inst. & Epochs & Binary status & Previous SpT & New SpT \\
\hline
WR\,20b & UVES  &   8    & candidate & WN6ha        & WN6ha   \\
MSP\,18 & UVES  &   8    & candidate & O5-5.5\,V-III((f)) & O5-5.5\,V-III((f))\\
MSP\,44 & UVES  &   8    & ecliping  & O9.5\,V      & B1\,V + PMS? \\
MSP\,96 & UVES  &   8    & eclipsing & OB           & B1: + B1: \\
MSP\,167& GIRAFFE-ARGUS& 3 & unknown & O6\,III      & earlier than O6 \\
MSP\,171& GIRAFFE-IFU & 5 & unknown  & O4-5         & O4-5 \\
MSP\,182& GIRAFFE-IFU & 5 & unknown  & O4           & O4 \\
MSP\,183& GIRAFFE-ARGUS& 3 & unknown & O3\,V((f))   & O3\,V((f)) \\
MSP\,199& GIRAFFE-ARGUS& 3 & unknown & O3-4\,V      & O3-4\,V \\
MSP\,203+444& GIRAFFE-IFU & 5 & unknown & O6\,V-III    & O4-5\,V \\ 
MSP\,223& UVES  &   8    & eclipsing?  & OB           & O7-8 \\
A       & GIRAFFE-ARGUS& 3 & unknown   & --           & O8\,V \\
B       & GIRAFFE-ARGUS& 3 & unknown   & --           & O6-7 \\
C       & GIRAFFE-ARGUS& 3 & unknown   & --           & O7\,V \\
D       & GIRAFFE-ARGUS& 3 & unknown   & --           & O9.5: \\
E       & GIRAFFE-ARGUS& 3 & unknown   & --           & O6-7:\\
\hline
\end{tabular}
\tablefoot{The first column lists the designation of the star, the second and third columns yield the instrument and the number of epochs for each target. The last three columns provide the binary status and spectral types prior to our study and the current spectral type (determined in this paper).}
\end{center}
\end{table*}

Determining the correct distance of this cluster is a fundamental issue: if the distance were confirmed to be 8.0\,kpc, Westerlund\,2 might be one of the most massive clusters of our Galaxy (along with NGC\,3603, Cyg\,OB2, and Westerlund\,1). Otherwise, if the distance were established to be much shorter than 8.0\,kpc, WR\,20a would be left in relative isolation, thereby challenging the theories of competitive accretion (Bonnell, Vine \& Bate \cite{Bonnell}) and the link between the mass of a cluster and the mass of its most massive member (Weidner et al.\ \cite{Weidner}). The distance is also of fundamental importance to understand the link with the TeV source HESS J1023$-575$ (Aharonian et al.\ \cite{Aharonian}), which is spatially consistent with the bright {\it Fermi}-LAT pulsar PSR J1023.0$-$5746 (Abdo et al.\ \cite{Abdo}, Ackermann et al.\ \cite{Ackermann}). At much longer wavelengths, a peculiar jet as well as arc-like molecular ($^{12}$CO) emission feature were reported by Fukui et al.\ (\cite{Fukui}). The arc is well centred on the TeV source HESS\,J1023$-$575, and Fukui et al.\ (\cite{Fukui}) argue that both the TeV and GeV $\gamma$-ray sources and the molecular features are likely to be associated with the same high-energy event. These authors accordingly suggest that these features could be due to an anisotropic supernova explosion of one of the most massive members of Westerlund\,2 or to remnant objects such as a pulsar wind nebula or a microquasar. On the basis of the detection of significant emission during the off-pulse intervals with {\it Fermi}-LAT, Ackermann et al.\ (\cite{Ackermann}) argued in favour of the pulsar wind nebula scenario. In either case, the association of such a remnant with Westerlund\,2 would imply that the most massive members of the cluster have already gone supernova, assuming star formation in this region was coeval. It must be stressed however that the preferred distance of PSR J1023.0$-$5746 is 2.4\,kpc (Ackermann et al.\ \cite{Ackermann}). Therefore, if the distance of Westerlund\,2 were established to be much larger than 2.5\,kpc, this would cast serious doubt on a connection between the pulsar and the cluster. 

In this paper, we present the results of a spectroscopic follow-up campaign of the eclipsing binaries and the stars in the core of Westerlund\,2. This campaign was specifically designed to address the questions outlined above. Throughout this work, we use the naming convention introduced by Moffat et al.\ (\cite{MSP}), except for objects that are discussed here for the first time. This paper is organized as follows. In Sect.\,\ref{observations}, we provide a brief overview of our observing campaign and the data reduction. Section\,\ref{results} presents our results concerning spectral classification, variability, and multiplicity for each star. The implications of these results are finally discussed in Sect.\,\ref{discussion}.

\begin{figure}[h!]
%[t!hb]
\begin{center}
\resizebox{8.5cm}{!}{\includegraphics{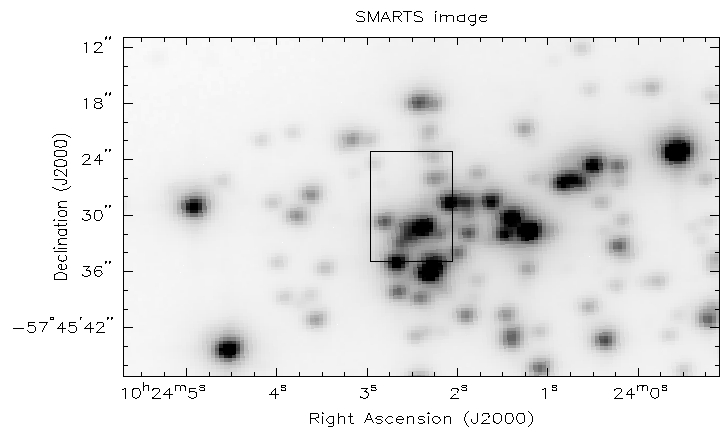}}

\resizebox{8.5cm}{!}{\includegraphics[angle=-90]{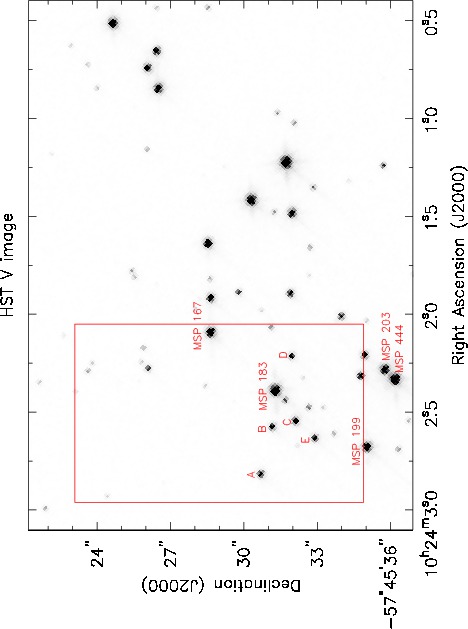}}
\end{center}
\caption{Illustration of the position of the ARGUS unit (box) on the core of Westerlund\,2. The top image uses our combined $B$ and $V$ image from Paper I (obtained with the SMARTS 1.3\,m telescope at CTIO), whilst the bottom image uses an archive WFPC2 HST (F555W) image. The various objects analysed in the ARGUS data are labeled in the bottom figure.\label{argus}}
\end{figure}
\section{Observations \label{observations}}
We obtained eight observations of the newly discovered eclipsing binaries in Westerlund\,2 (Paper I), as well as several O-star binary candidates with the FLAMES multi-object spectrograph on the Very Large Telescope (VLT) UT2 (see Table\,\ref{journal}). The observations were performed in service mode over an interval of one week in March 2008. Each observation lasted for an integration time of 2775\,s and the observing conditions were good with a seeing of better than 1.1\,arcsec. FLAMES offers the possibility to simultaneously acquire high-resolution spectra of up to eight targets with the UVES spectrograph and medium-resolution spectra of some additional targets with the GIRAFFE spectrograph. 

\begin{table*}[t!]
\caption{Radial velocities of prominent interstellar absorption and nebular emission lines in the FLAMES-UVES spectra of our targets. \label{ISM}}
\begin{center}
\begin{tabular}{l c c c c c}
\hline
Line & WR\,20b & MSP\,18 & MSP\,44 & MSP\,96 & MSP\,223 \\
& (km\,s$^{-1}$) & (km\,s$^{-1}$) &(km\,s$^{-1}$) &(km\,s$^{-1}$) &(km\,s$^{-1}$) \\
\hline
Na\,{\sc i} $\lambda$\,5890       & $3.4 \pm 0.3$ & $2.8 \pm 0.3$& $2.8 \pm 0.5$& $2.2 \pm 0.2$& $2.3 \pm 0.4$\\
Na\,{\sc i} $\lambda$\,5896       & $3.7 \pm 0.2$ & $2.0 \pm 0.5$& $2.9 \pm 0.5$& $1.8 \pm 0.3$& $3.2 \pm 0.2$\\
$[$O\,{\sc iii}$]$ $\lambda$\,4959& $17.1 \pm 7.6$&$-3.7 \pm 1.0$&$28.3 \pm 1.7$&$22.3 \pm 0.3$& $7.4 \pm 2.6$\\
$[$O\,{\sc iii}$]$ $\lambda$\,5007& $18.6 \pm 1.7$&$-3.1 \pm 1.5$&$29.7 \pm 1.2$&$22.6 \pm 0.8$& $8.2 \pm 0.9$\\
\hline
\end{tabular}
\tablefoot{The error bars correspond to the 1-$\sigma$ dispersion in our measurements}
\end{center}
\end{table*}

Our campaign was divided into two parts (run A and B) with two different instrumental set-ups for the GIRAFFE spectrograph. During all our observations, five well-isolated targets (WR\,20b, MSP\,18, as well as the three eclipsing systems MSP\,44, 96, and 223) were observed with FLAMES-UVES fibres. The UVES echelle spectrograph was operated in the standard setup Red 580, providing coverage of the wavelength domains 4842 -- 5728\,\AA\ and 5843 -- 6800\,\AA\ with a resolving power of 47000. During run A (five observations), the GIRAFFE spectrograph was fed by three integral field units (IFU), allowing us to monitor three additional O-type stars (MSP\,171, 182, and 203). Whilst MSP\,182 and 171 are well-isolated targets, MSP\,203 lies in a more crowded region near the cluster core and its spectrum is heavily blended with that of MSP\,444 (see Moffat et al.\ \cite{MSP}). During run B (three observations), GIRAFFE was used with the ARGUS system. ARGUS consists of an array of $14 \times 22$ microlenses offering integral field spectroscopy over a field of $11.8 \times 7.3$\,arcsec with a sampling of 0.52\,\arcsec. The field covered by ARGUS was oriented north-east and was centered at the position $\alpha = $10:24:02.5, $\delta = -$57:45:24.0 (see Fig.\,\ref{argus}) near MSP\,183 and MSP\,167\footnote{The coordinates quoted in this paper are those from the archive WFPC2 HST image (based on GSC). They are shifted by $-0.027$\,s in RA and $-0.69$'' in Dec with respect to the 2MASS system used in Paper I.}. For both runs (A and B), the GIRAFFE spectrograph was used in the low-resolution standard setup LR04 in third order covering the domain 5015 -- 5815\,\AA\ with a measured resolving power of 9500.

All the data were reduced with the dedicated pipelines (versions 4.2.1 for UVES, 2.5.1 for GIRAFFE-IFU and 2.8.1 for GIRAFFE-ARGUS). The signal-to-noise ratio (S/N) of the individual FLAMES-UVES spectra reached between about 50 for the faintest stars and around 250 for WR\,20b and MSP\,18. For the three FLAMES-GIRAFFE spectra, the S/N was around 150 for MSP\,171 and 182 and near 300 for MSP\,203 + MSP\,444. For the ARGUS data, the wavelength solution was obtained in two iterations yielding an rms in the calibration of 2.7\,km\,s$^{-1}$. The spectra were extracted using the SUM option.

Owing to the high extinction towards Westerlund\,2, the spectra of our targets are very rich in interstellar features. Besides the strong Na\,{\sc i} $\lambda\lambda$\,5890, 5896 absorption lines, we observe a large number of diffuse interstellar bands (DIBs, see Herbig \cite{Herbig}). In addition, the spectra display a number of nebular emission lines from the RCW\,49 complex, such as $[$O\,{\sc iii}$]$ $\lambda\lambda$\,4959, 5007 as well as H\,{\sc i} and He\,{\sc i}. 

As can be seen from the results quoted in Table\,\ref{ISM}, the interstellar absorption lines are consistently found to indicate radial velocities (RVs) of around 2 -- 3\,km\,s$^{-1}$ for each star, whilst the radial velocities of the nebular emission lines span a range from  $-3.7$ to $+29.7$\,km\,s$^{-1}$, implying that the RCW\,49 nebula has a rather complex velocity field. 

\begin{figure}[htb!]
\resizebox{8.5cm}{!}{\includegraphics{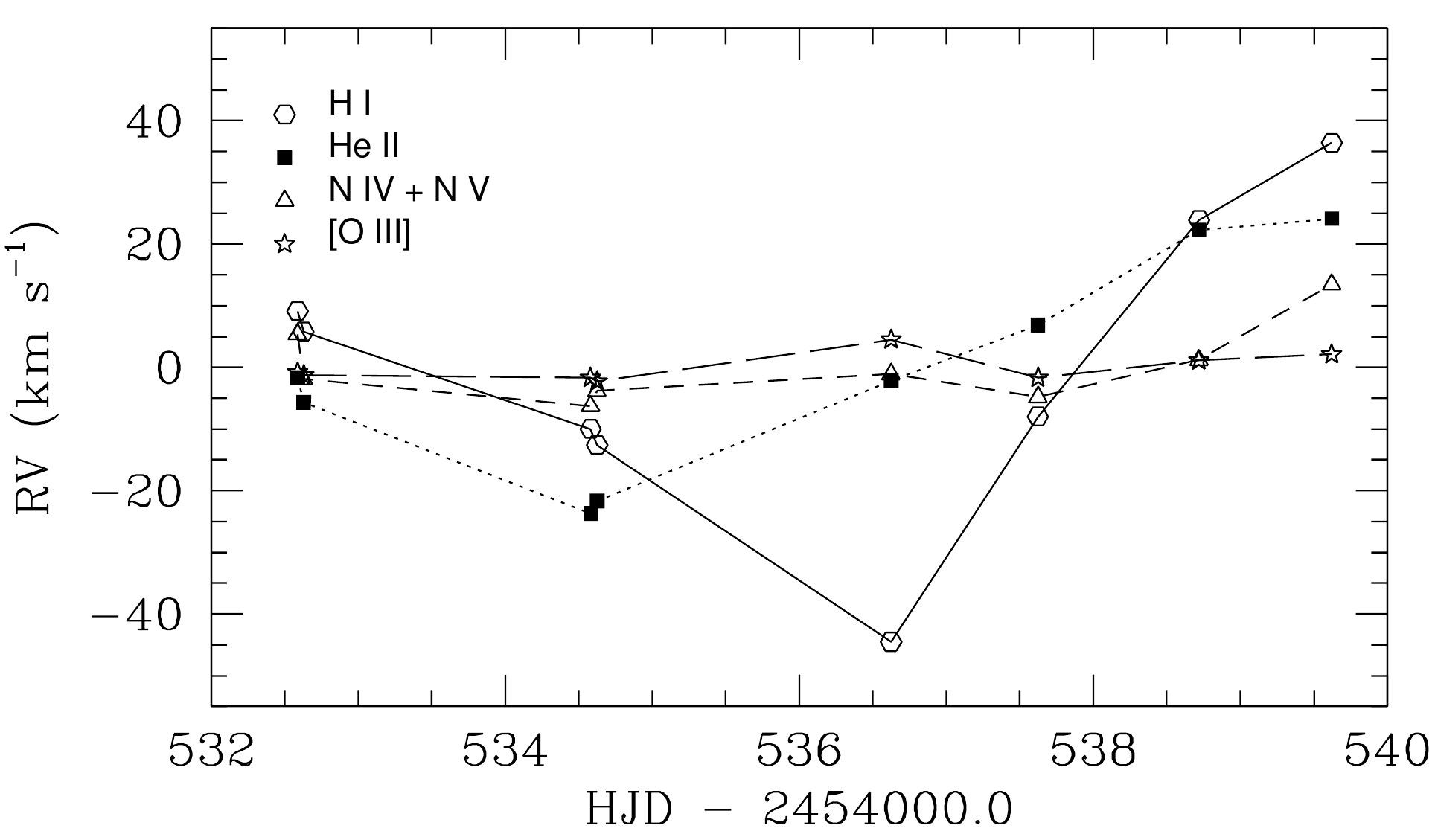}}
\caption{Radial velocity variations about the mean evaluated from the five measurements for several sets of lines in the spectrum of WR\,20b (see text). The RV measurements of the narrow nebular $[$O\,{\sc iii}$]$ $\lambda$\,5007 emission line are shown for comparison.\label{WR20bRV}}
\end{figure}

\section{Results \label{results}}
\subsection{WR\,20b}
WR\,20b was reported as an H$\alpha$ emission star by Th\'e (\cite{The}, his object THA 35-II-37). This result was confirmed by Schwartz, Persson \& Hamann (\cite{SPH}), and the star was first classified as a WN7 Wolf-Rayet object by Shara et al.\ (\cite{SMSP}). This classification was slightly revised to WN7:h, i.e.\ a WN7 star with a definite signature of hydrogen in its spectrum, by Shara et al.\ (\cite{Shara}), whilst van der Hucht (\cite{vdH}) proposed instead a spectral type WN6ha, i.e.\ a WN6 star with hydrogen emission lines and absorption lines belonging to the WN6 star. Subsequent radio observations revealed the presence of a wind-blown shell around WR\,20b (Whiteoak \& Uchida \cite{WU}). Whilst the mean photometric properties of WR\,20b do not support the presence of a luminous companion, the X-ray spectrum of the star is remarkably hard, a feature that is often associated with interacting wind binaries (Naz\'e et al.\ \cite{NRM}). 
\begin{figure*}[t!hb]
\begin{center}
\resizebox{12.0cm}{!}{\includegraphics{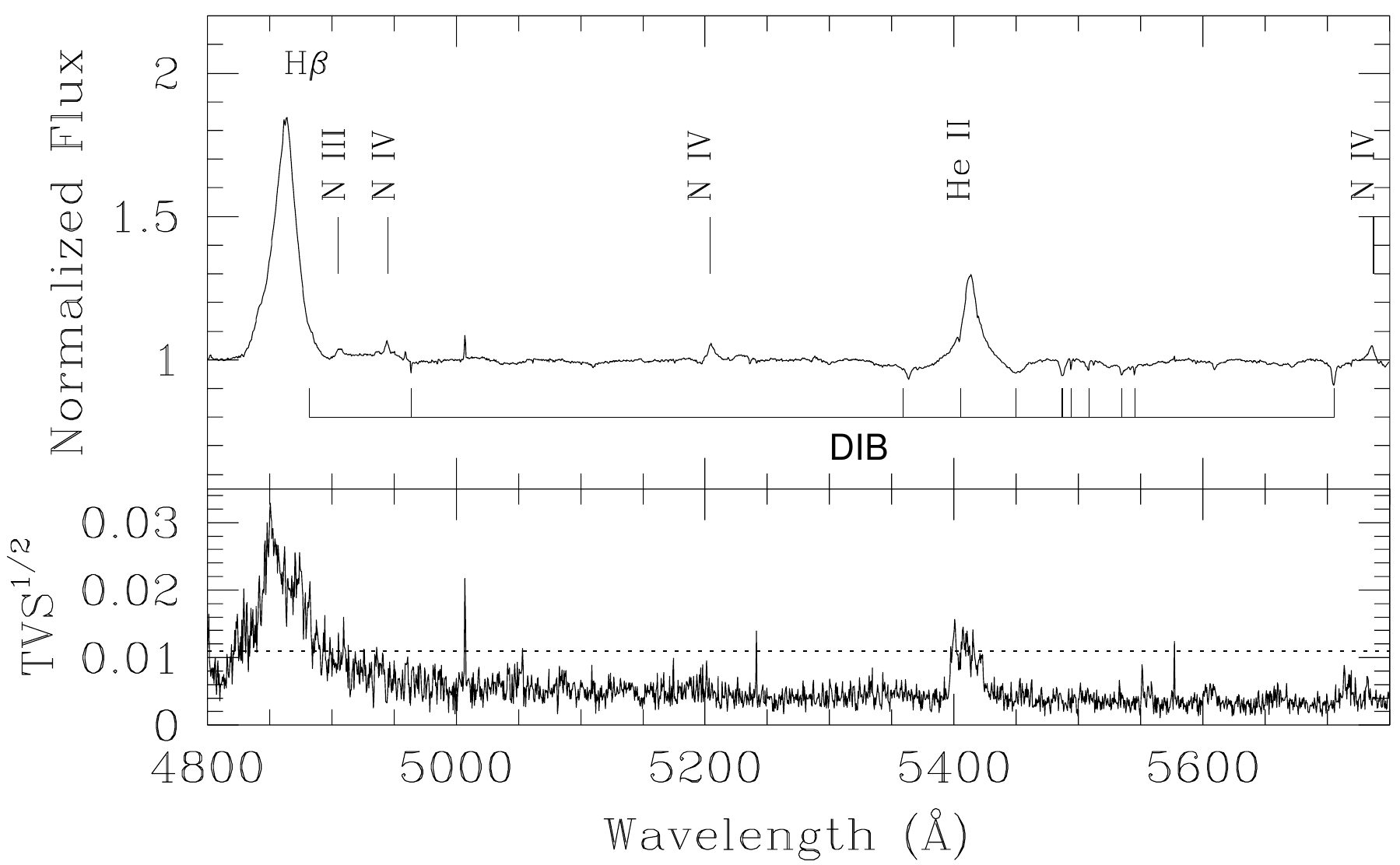}}

\resizebox{12.0cm}{!}{\includegraphics{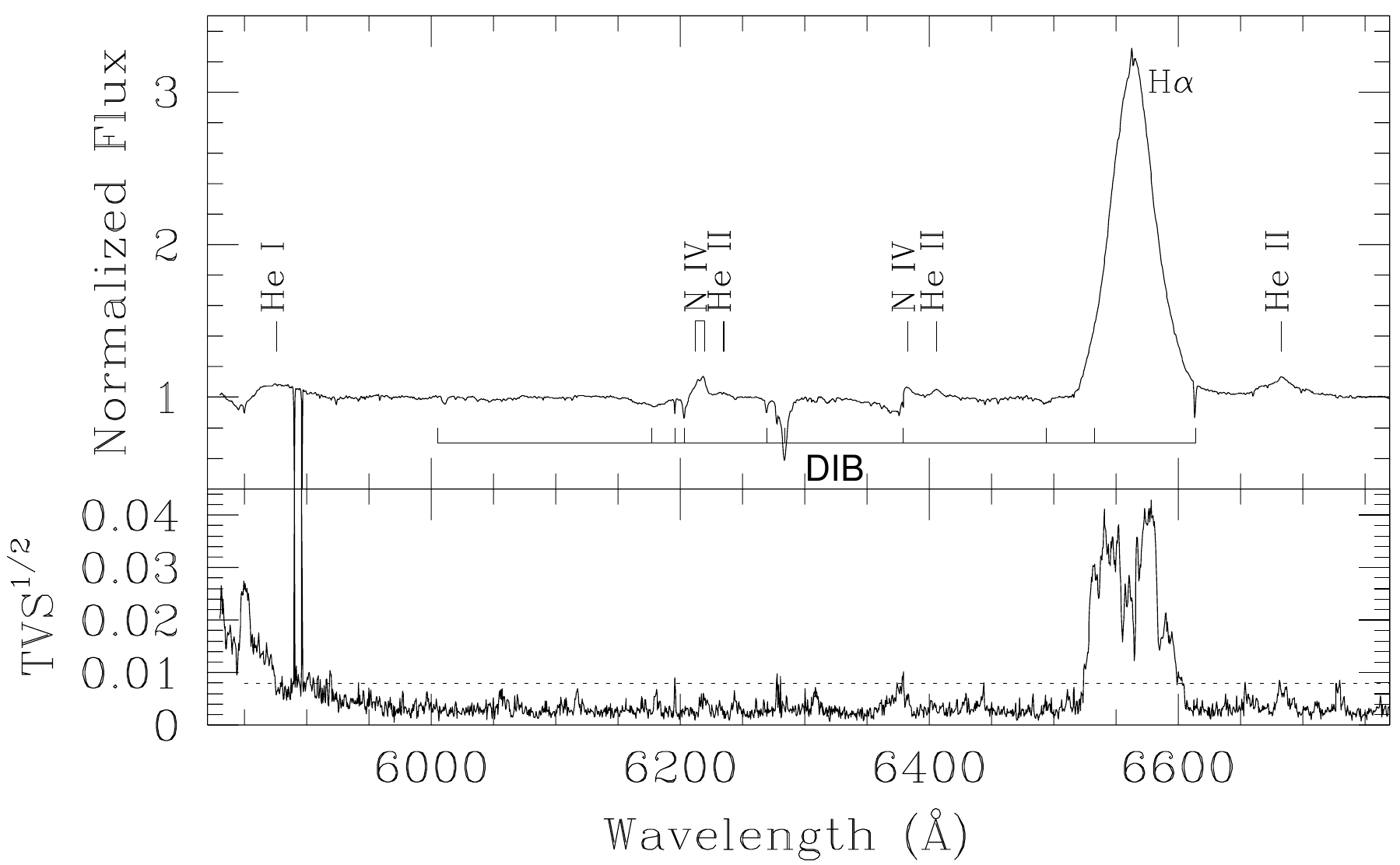}}
\end{center}
\caption{Mean spectrum and temporal variance spectrum as computed from our time series of FLAMES-UVES spectra of WR\,20b. The dashed lines yield the 99\% significance level of the TVS$^{1/2}$.\label{WR20bTVS}}
\end{figure*}

To address the issue of multiplicity, we measured the RVs of a number of stellar emission lines (H$\beta$, H$\alpha$, He\,{\sc ii} $\lambda\lambda$\,5412, 6406, 6683, N\,{\sc iv} $\lambda\lambda$\,5204, 5737, 6383, and N\,{\sc v} $\lambda$\,4945, see Table\,\ref{RV20b}), as well as interstellar and nebular features such as the diffuse interstellar bands (DIBs) at $\lambda\lambda$\,5363, 5705, 5850, 6010, 6614 (see Herbig \cite{Herbig}), the Na\,{\sc i} $\lambda\lambda$\,5890, 5896 doublet, and the $[$O\,{\sc iii}$]$ $\lambda$\,5007 line. The interstellar and nebular features allow us to check the stability of the wavelength calibration: their $1-\sigma$ RV dispersion is 0.3\,km\,s$^{-1}$ for the narrow Na\,{\sc i} absorptions, indicating a very stable calibration. For the broad DIBs, the dispersion is larger, of the order of 4\,km\,s$^{-1}$. For the majority of the stellar emission lines, the RV dispersions are significantly larger than these values (24.8, 17.7, and 6.5\,km\,s$^{-1}$ for the H\,{\sc i}, He\,{\sc ii} and N\,{\sc iv} + {\sc v} lines, respectively). However, as shown in Fig.\,\ref{WR20bRV}, the variations in the different sets of lines are uncorrelated. 

\begin{figure}[htb]
\resizebox{8.5cm}{!}{\includegraphics{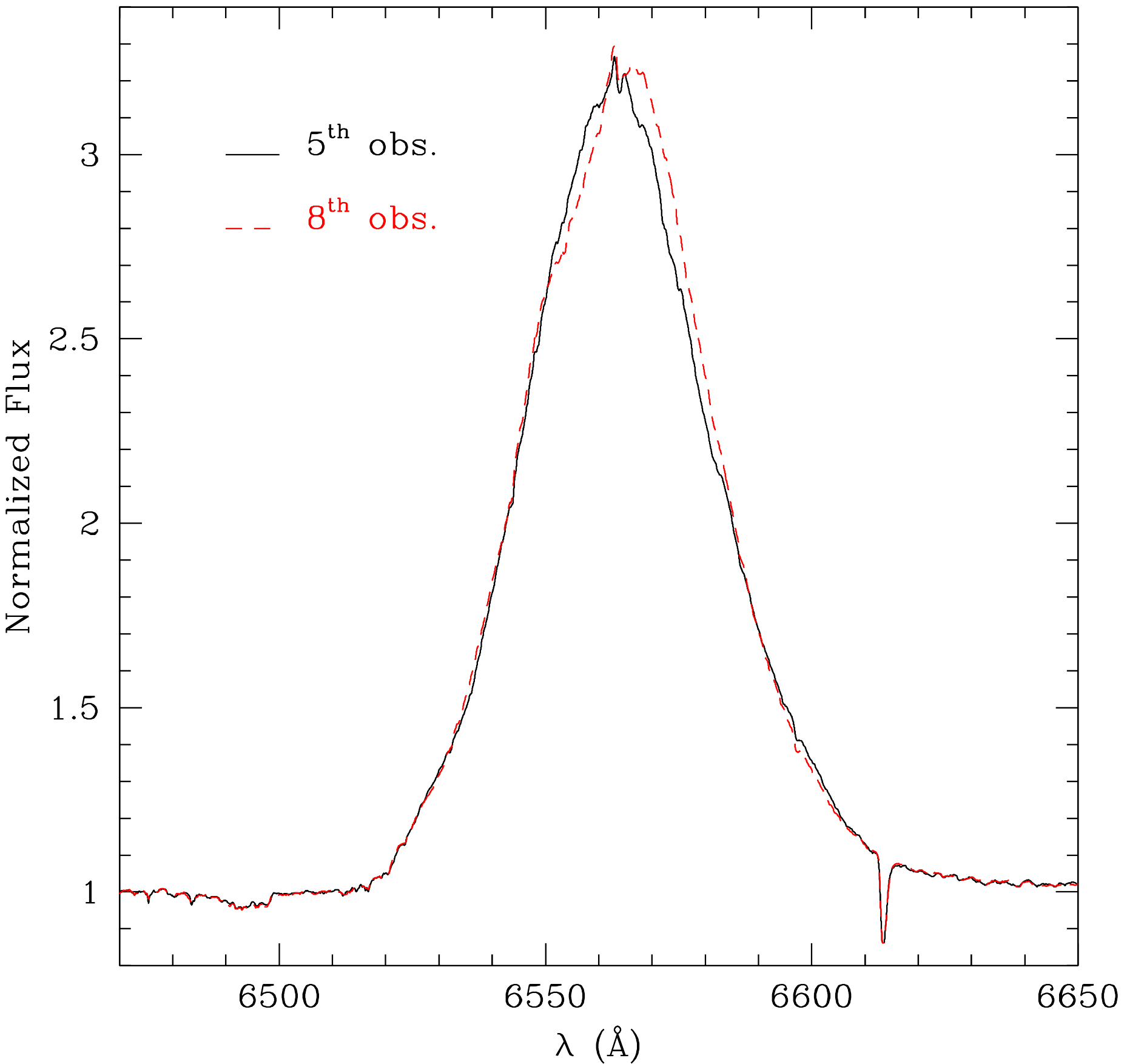}}
\caption{Illustration of the H$\alpha$ line profile in the spectrum of WR\,20b for those two observations corresponding to the maximum difference in apparent RV. The RV variations are clearly due to the top part of the line profile only. \label{WR20bHalpha}}
\end{figure}

To quantify the spectral variability of the star, we computed the temporal variance spectrum (TVS, see Fullerton, Gies \& Bolton \cite{FGB}) of our time series. This method provides a quantitative assessment of the significance of the spectral line profile variability properly accounting for the noise level of the individual spectra. The results are shown in Fig.\,\ref{WR20bTVS}. Significant variations are detected in the strongest emission lines, but not in weaker ones such as the N\,{\sc iv} $\lambda\lambda$\,6212-20 blend. The TVS of the strongest lines is quite complex, with multiple (up to four) peaks, indicating that the RV variability of these lines reflects their line profile variability. A more detailed inspection of the profiles of the hydrogen lines in different observations (see Fig.\,\ref{WR20bHalpha}) actually indicates that the apparent RV variations are mainly due to changes in the upper part of the line profiles.  
The results presented in this section suggest that WR\,20b is indeed a spectroscopically variable object, but this variability is more likely to be related to (large-scale) structures in the stellar wind, than caused by genuine RV variations of the star. Therefore, we conclude that there is at present no firm evidence that WR\,20b is a short period (less than two week period) spectroscopic binary system (see also the discussion in Sect.\,\ref{discussion}).

\begin{figure*}[h!tb]
\begin{center}
\resizebox{12.0cm}{!}{\includegraphics{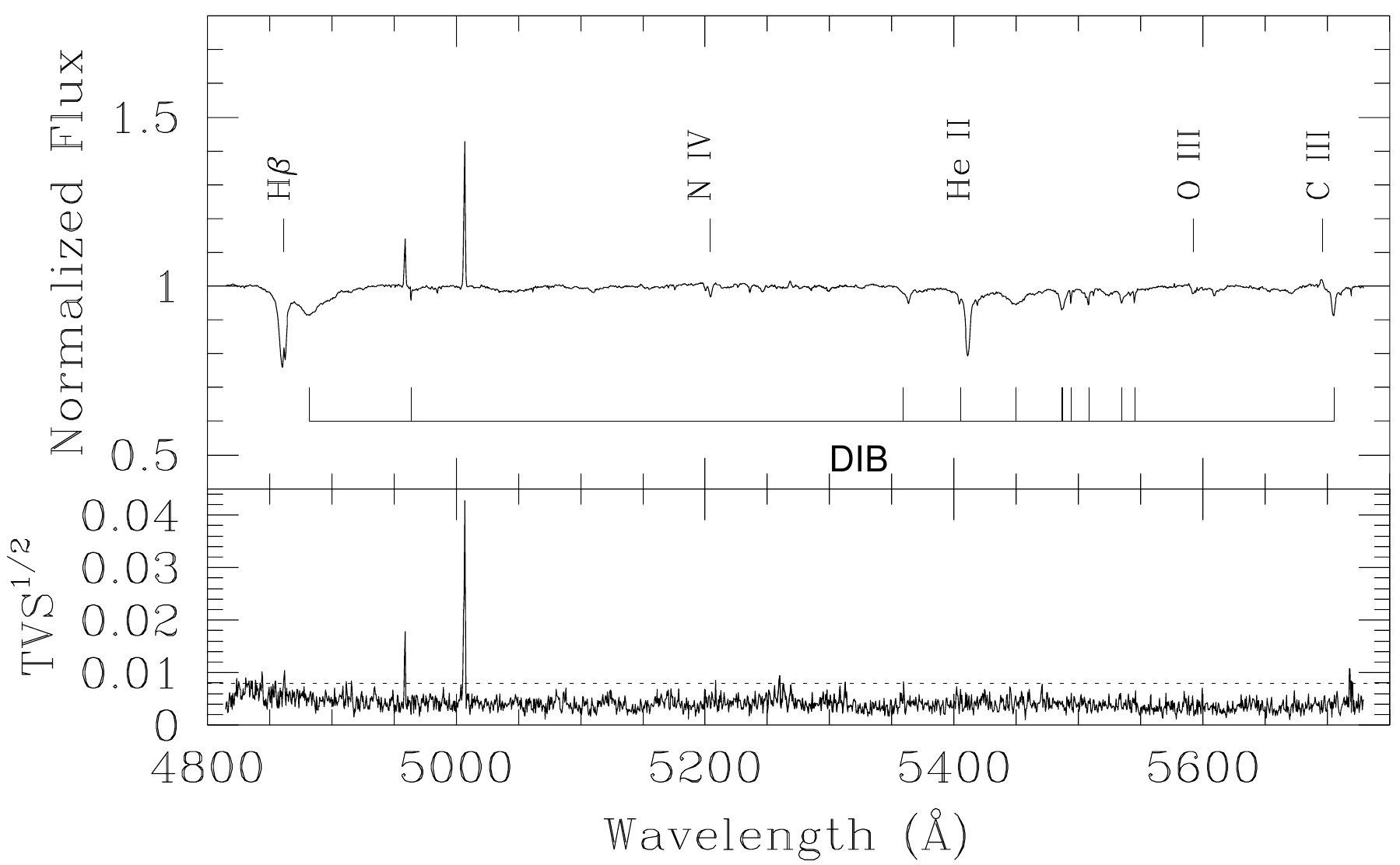}}

\resizebox{12.0cm}{!}{\includegraphics{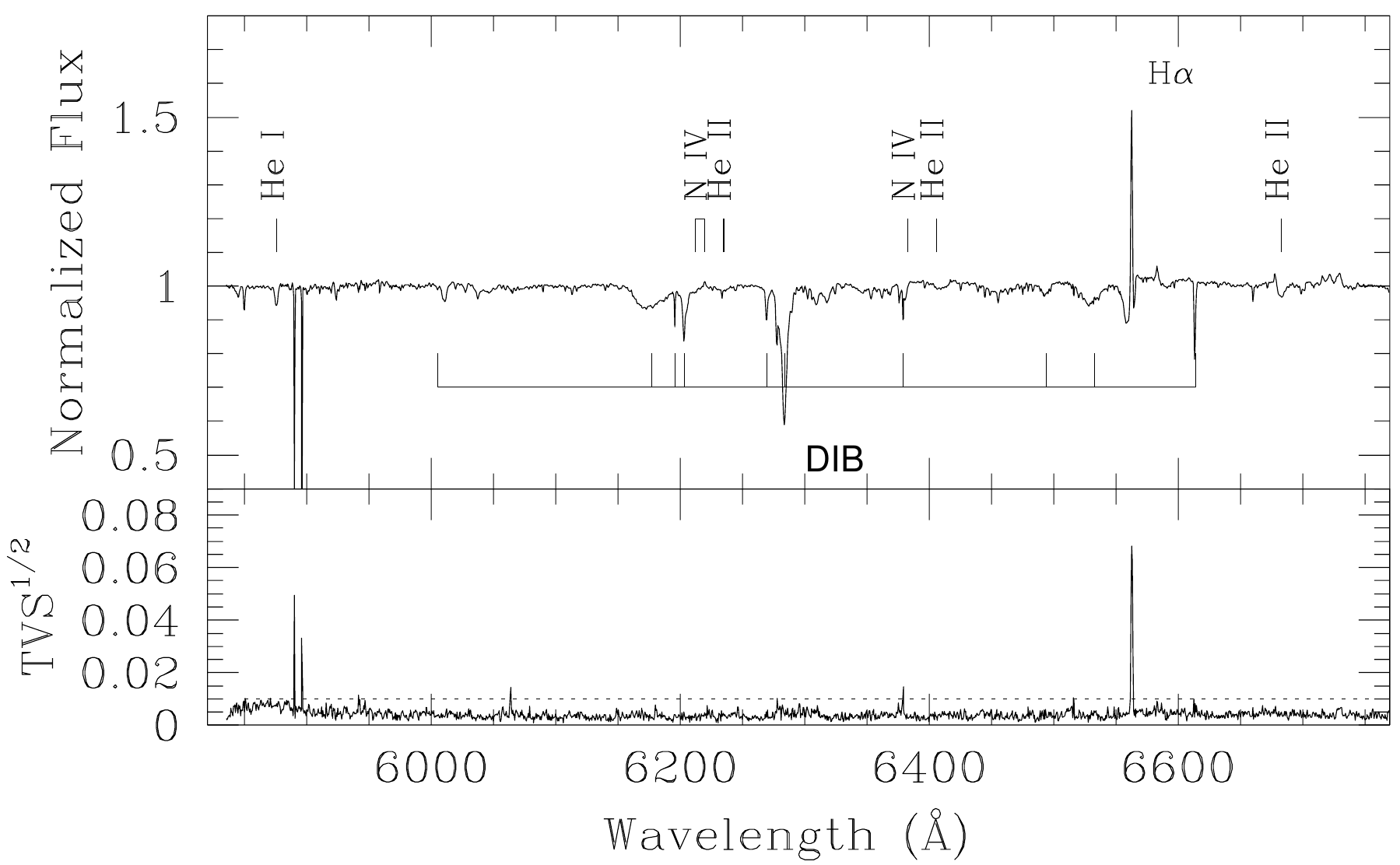}}
\end{center}
\caption{Same as Fig.\,\ref{WR20bTVS}, but for MSP\,18. \label{MSP18TVS}}
\end{figure*}

Following the classification criteria of Smith, Shara \& Moffat (\cite{SSM}), our spectra correspond most closely to a WN5-6 spectral type (ratio of the He\,{\sc ii} $\lambda$\,5412 to He\,{\sc i} $\lambda$\,5876 intensities above continuum = 3.6) with narrow lines (equivalent width, EW, of He\,{\sc ii} $\lambda$\,5412 = $-5.9$\,\AA\ $> -40$\,\AA) and a definite signature of hydrogen emission revealed by the oscillation of the Pickering decrement (compare the intensities of H$\alpha$, He\,{\sc ii} $\lambda$\,5412, and H$\beta$ on Fig\,\ref{WR20bTVS}). The only obvious stellar absorption feature in our spectra is the P-Cygni-type absorption trough of the He\,{\sc i} $\lambda$\,5876 line, which is obviously associated with the Wolf-Rayet star. Therefore, our data are consistent with the WN6ha classification given by van der Hucht (\cite{vdH}).

\subsection{MSP\,18}
On the basis of low-resolution spectroscopy, Moffat et al.\ (\cite{MSP}) classified MSP\,18 as an O7:V star. This classification was revised to O4\,V((f)) by Uzpen et al.\ (\cite{Uzpen}) and more recently to O5-5.5\,V-III((f)) based on higher resolution spectra (Paper I). Both its optical brightness (Paper I) and X-ray emission properties (brightness, spectral hardness and slight flux variability; see Naz\'e et al.\,\cite{NRM}) suggest that this star could be an interacting-wind binary system. 

We fitted Gaussian line profiles to measure the RVs of a number of stellar absorption lines (H$\beta$, H$\alpha$, He\,{\sc i} $\lambda$\,5876, He\,{\sc ii} $\lambda\lambda$\,5412, 6683), as well as two faint emission lines (C\,{\sc iii} $\lambda\lambda$\,5696, 6729-6731; Walborn \cite{Walborn}). The results are listed in Table\,\ref{RVMSP18}. We also measured the same interstellar and nebular features as for WR\,20b as well as the nebular H$\beta$ and H$\alpha$ emission lines. The $1-\sigma$ RV dispersions of the stellar absorption lines are between 1.8 and 5.9\,km\,s$^{-1}$, which is quite comparable to the corresponding values of the prominent interstellar features. The stellar emission lines have slightly larger RV dispersions, but these emission lines are actually extremely weak (normalized intensity of 0.02 above the continuum for C\,{\sc iii} $\lambda$\,5696) and their RVs are thus subject to larger uncertainties. In addition, the TVS does not reveal any significant line profile variability. We thus conclude that MSP\,18 is unlikely to be a short period (shorter than two week) spectroscopic binary (see also the discussion in Sect.\,\ref{discussion}).

\subsection{MSP\,96}
The star MSP\,96 was found to be an OB-type eclipsing binary system with a period of $(1.0728 \pm 0.0006)$\,days with probably at least one component close to filling its Roche lobe (Paper I). Our spectra do not reveal any trace of the He\,{\sc ii} $\lambda$\,5412 line, thus they are most consistent with a spectral type later than B0.2 (Walborn \cite{NRW}). As a result of the severe contamination of the spectra by strong and asymmetric nebular emission lines in the hydrogen and neutral helium transitions, measuring the stellar components in the spectra of MSP\,96 is quite a challenging task. 

The stellar spectral lines are broad, as expected for rapidly rotating stars in a contact binary system. Using a multi-Gaussian fitting technique, we attempted to deblend the H$\alpha$, He\,{\sc i} $\lambda\lambda$\,5871, 6678 lines, accounting for the presence of a strong nebular emission line. Our most reliable results were obtained for the He\,{\sc i} $\lambda$\,6678 line and the corresponding RVs are listed in Table\,\ref{RV96}. Unfortunately, the uncertainties on these RVs are quite large, of the order of 20\,km\,s$^{-1}$.
The He\,{\sc i} lines are essentially identical in strength for the spectra of both stars, whilst the H$\alpha$ line appears to be up to four times stronger in the spectrum of the star that has negative velocities during most of our observations. We caution however that this latter result is probably an overestimate because of a bias resulting from the presence of the strong nebular emission line. 

\begin{table}
\caption{Radial velocities of MSP\,96.\label{RV96}}
\begin{center}
\begin{tabular}{c c c c c}
\hline
Date & $\phi_p$ & $\phi_{RV}$ & Primary & Secondary \\
HJD-2454000 & & & (km\,s$^{-1}$) & (km\,s$^{-1}$) \\
\hline
532.587 & 0.044 & 0.058 &        &   154  \\
532.631 & 0.085 & 0.099 & $-182$ &   159  \\
534.581 & 0.903 & 0.917 &   239  & $-113$ \\
534.628 & 0.946 & 0.960 &   112  &  $-80$ \\
536.624 & 0.807 & 0.821 &   289  & $-230$ \\
537.626 & 0.741 & 0.755 &   374  & $-259$ \\
538.719 & 0.760 & 0.774 &   352  & $-219$ \\
539.622 & 0.602 & 0.616 &   195  & $-167$ \\
\hline
\end{tabular}
\tablefoot{The RVs are inferred from the multi-Gaussian deblending of the He\,{\sc i} $\lambda$\,6678 line accounting for a nebular emission line. The second and third columns yield the orbital phases computed according to the photometric and spectroscopic ephemerides, respectively.}
\end{center}
\end{table}

In our data, the two components are most clearly deblended for the fifth, sixth, and seventh observations. Extrapolating the light curve ephemerides of Paper I to the epoch of our spectroscopic observations, we find that these observations should correspond to orbital phases near $\phi_p = 0.75$ (where $\phi_p = 0.0$ corresponds to the primary eclipse, i.e.\ the brighter star being eclipsed). Although our spectroscopic data are taken 1162 days after $T_0$ of Paper I and despite the short orbital period, the orbital solution below agrees with the photometric ephemerides. Indeed, the shift in phase between the photometric and spectroscopic solution is found to be 0.014. 

We note that the star with the stronger H$\alpha$ line is seen blueshifted at phases near $\phi_p = 0.75$, hence that this line belongs to the intrinsically fainter component (i.e.\ the photometric secondary) of the system. From these considerations, we conclude that the photometric secondary must be intrinsically somewhat cooler than the primary, but should have a somewhat larger radius. The helium line intensity ratio might then be close to unity, although the primary star (i.e.\ the one that is eclipsed at phase 0.0) would be somewhat hotter, thus explaining the deeper primary eclipse. Alternatively, the star with the negative RVs near $\phi_p = 0.75$ could be the primary star, but this would imply a shift in phase between the photometric and spectroscopic ephemerides of 0.486. This seems {\it a priori} rather unlikely, but cannot entirely be ruled out given the error in the orbital period and the large number of orbital cycles between our photometric and spectroscopic campaigns. We briefly consider the implications of this alternative assumption below.  
\subsubsection{Orbital solution\label{orbit96}}
To establish the orbital solution of MSP\,96 from the RV data in Table\,\ref{RV96}, we started by considering the SB1 solution for the photometric secondary star. For this purpose, we used the Li\`ege Orbital Solution Package (LOSP) code (Sana \& Gosset \cite{SG}), which allowed us to derive $K_s = (270 \pm 22)$\,km\,s$^{-1}$, as well as $\gamma = (19 \pm 19)$\,km\,s$^{-1}$. We then considered the RVs of the photometric primary star with the value of $\gamma$ fixed to $19$\,km\,s$^{-1}$, which yielded $K_p = (337 \pm 78)$\,km\,s$^{-1}$. From these results, we then inferred a secondary/primary mass ratio of $1.25 \pm 0.27$, as well as $m_p\,\sin^3{i} = (11.1 \pm 3.0)\,M_{\odot}$ and $m_s\,\sin^3{i} = (13.8 \pm 6.6)\,M_{\odot}$. 
These minimum masses, along with the eclipsing nature of the system, which indicates a rather large inclination, suggest a spectral type near B1 for both components. This classification is also consistent with the overall spectral morphology, although the large number of interstellar features and the moderate S/N of our data prevent us from being more specific.
\subsubsection{Light curve solution}
Adopting the mass ratio inferred hereabove, we then analysed the photometric data of the system from Paper I using the {\sc nightfall}\footnote{ http://www.hs.uni-hamburg.de/DE/Ins/per/Wichmann/Nightfall.html} code, developed and maintained by R.\ Wichmann, M.\ Kuster and P.\ Risse. We set the temperature of the primary star to 25400\,K (appropriate for a B1\,V classification, Schmidt-Kaler \cite{SK}) and allowed the inclination, the Roche-lobe filling factors (hereafter {\it fill}$_p$ and {\it fill}$_s$), and the secondary temperature to vary. The best-fit model parameters were found to be $i = (76.8^{+0.1}_{-0.2})^{\circ}$, $T_s = 21550^{+120}_{-260}$\,K, {\it fill}$_p = 0.84 \pm 0.03$, and {\it fill}$_s = 0.85 \pm 0.02$. This inclination yields absolute masses and radii, respectively of $M_p = (12.0 \pm 3.3)\,M_{\odot}$, $M_s = (15.0 \pm 7.2)\,M_{\odot}$, and $R_p = (3.8 \pm 0.6)\,R_{\odot}$, $R_s = (4.3 \pm 0.6)\,R_{\odot}$. These radii are rather small for the masses, but we note that quite similar parameters were found for other early B-type main-sequence stars in eclipsing binaries (e.g.\ V\,578 Mon and DW\,Car, see Southworth \cite{Southworth} and references therein). That the more massive star (the secondary) would be the cooler and less luminous component of the system is surprising and indicates that this star would actually be more evolved. The radii of both stars suggest instead that they are rather unevolved main-sequence stars. Whilst a case A binary mass transfer could alter the properties of the components, we stress that the light curve solution relates to stars that do not fill up their Roche lobes. This issue would actually be resolved if we assumed a phase shift of 0.486 between the photometric and spectroscopic ephemerides (see below).  
\begin{figure}[htb]
\resizebox{8.5cm}{!}{\includegraphics{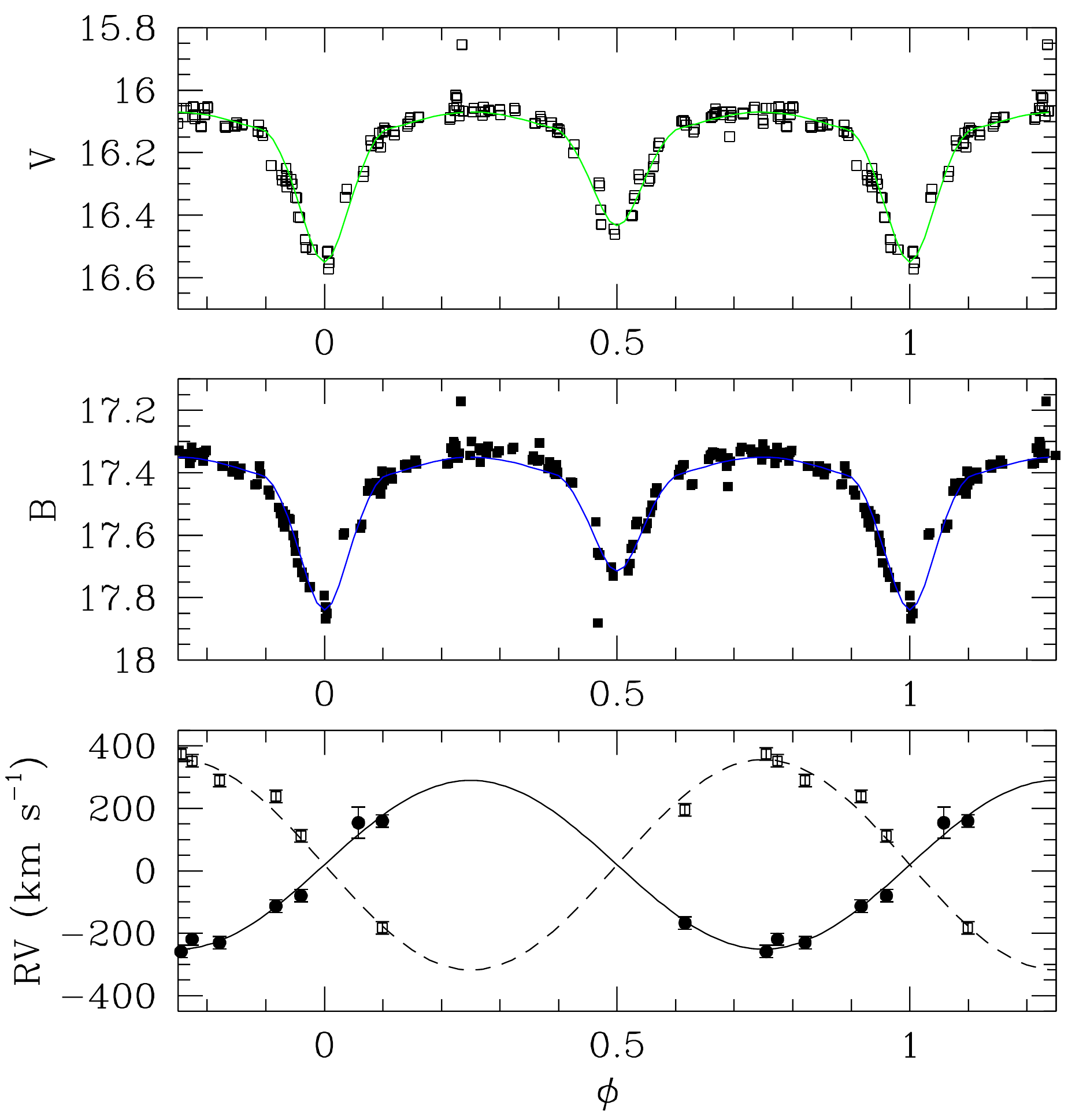}}
\caption{Combined solution of the light curve and the RV curve of the  MSP\,96 system (see text). The orbital phases are computed from the photometric ephemerides of Paper I.\label{MSP96sol}}
\end{figure}

Combining all the above parameters, we inferred a total bolometric luminosity of the binary of $8960\,L_{\odot}$. The uncertainty in this number is difficult to evaluate, because of the unknown uncertainty in the primary's spectral type. However, we estimated that the relative uncertainty is likely to be about 25\%. From $V = 16.17$ at phases outside the eclipses and $E(B-V) = 1.50$ as follows from Fig.\,7 of Paper\,I, we then inferred a distance of 6.6\,kpc again probably subject to a 25\% uncertainty.   

We also analysed the light curve assuming a shift of 0.486 in phase between the photometric and spectroscopic ephemerides. In this case, the mass ratio inferred in Sect.\,\ref{orbit96} would be inverted, $m_s/m_p = 0.80 \pm 0.17$, and the more massive star (which is now the primary with $m_p\,\sin³{i} = 13.8\,M_{\odot}$) would also be the brighter one. The corresponding best-fit photometric solution is essentially symmetric with respect to the parameters indicated above ($i = 77.2^{\circ}$, $T_p = 25400$\,K (fixed), $T_s = 21790$\,K, {\it fill}$_p = 0.85$, {\it fill}$_s = 0.84$) and yields a similar distance of $6.8$\,kpc. This solution corresponds to an unevolved B + B main-sequence binary system and avoids the problems described above.

\subsection{MSP\,44}
In Paper I, we classified MSP\,44 as an eclipsing binary with a period of 5.176\,days, noting that the aliasing period at 1.2395\,days yielded an almost equally strong peak in the periodogram. However, the rather narrow eclipses argue against the shorter orbital period and we thus favour the solution corresponding to the $(5.176 \pm 0.029)$\,day period. Extrapolating the photometric ephemerides from Paper I to the epoch of our FLAMES observations, we find the phases listed in the second column of Table\,\ref{tablemsp44}, where phase 0.0 corresponds to the primary eclipse. 

Whilst the primary eclipse has a depth of 0.6\,mag, the secondary eclipse was found to be much shallower ($< 0.1$\,mag). There are two possible interpretations of this situation: either the surface brightnesses of the two components of the system are very different or the binary orbit is sufficiently eccentric with a longitude of periastron (for the star being eclipsed at phase 0.0) near $90^{\circ}$\footnote{If the eccentricity were sufficiently large and the longitude of periastron close to $90^{\circ}$, one could expect to observe a difference in the durations of the primary and secondary eclipses, although the amplitude of this effect strongly depends on a number of parameters (orbital eccentricity and inclination, filling factors,...) and is sometimes even absent. Within the accuracy of our data, no such effect is present in the light curve of MSP\,44.}. In addition, the light curve exhibits some slow modulation, especially around secondary minimum that we tentatively attributed to spots on one of the stars (Paper I). 

\begin{table*}
\caption{Radial velocities of MSP\,44. \label{tablemsp44}}
\begin{center}
\begin{tabular}{c c c c c c c}
\hline
HJD         & $\phi_p$ & $\phi_{RV}$ & \multicolumn{3}{c}{He\,{\sc i} $\lambda$\,5876} & \multicolumn{1}{c}{He\,{\sc i} $\lambda$\,4921} \\
-2454000    &          &             & RV$_1$ & RV$_b$ & RV$_r$ & RV$_1$\\
            &          &             &(km\,s$^{-1}$) & (km\,s$^{-1}$) & (km\,s$^{-1}$) & (km\,s$^{-1}$)\\
\hline
532.587 & 0.273 & 0.157 & 17.3 & $-285$ &     &  5.3 \\
532.631 & 0.281 & 0.166 & 26.4 & $-269$ &     & 25.6\\
534.581 & 0.658 & 0.542 & 45.9 & $-252$ & 309 & 59.8\\
534.628 & 0.667 & 0.551 & 36.6 & $-267$ & 287 & 66.5\\
536.624 & 0.053 & 0.937 &      & $-246$ &     & 46.5\\
537.626 & 0.247 & 0.131 & 27.0 & $-273$ &     & 12.1\\
538.719 & 0.458 & 0.342 &      &        &     & 29.6\\
539.622 & 0.632 & 0.516 & 45.1 & $-264$ & 315 & 27.6\\
\hline
\end{tabular}
\tablefoot{The second and third columns yield the orbital phases computed according to the photometric and spectroscopic ephemerides respectively. RV$_1$, RV$_b$, and RV$_r$ stand for the RVs of the main component of the He\,{\sc i} $\lambda$\,5876 line and the blue and red components of this line, respectively (see text).}
\end{center}
\end{table*}

The absence of any He\,{\sc ii} line in the spectrum of MSP\,44 clearly indicates a spectral type later than B0.2 (Walborn \cite{NRW}) for the combined spectrum of MSP\,44. This is slightly at odds with the O9.5V: photometric spectral classification of this star (Naz\'e et al.\,\cite{NRM}). The likely presence of a very weak Si\,{\sc iii} $\lambda$\,5740 feature and possibly of Si\,{\sc ii} $\lambda$\,6347 suggest a spectral type near B1 with an uncertainty of at least one subtype. Given the probably rather large brightness ratio of the primary to the secondary, we can actually adopt this spectral type as the one of the primary component. The B1 spectral classification is also surprising in view of the rather strong and hard X-ray emission of MSP\,44 measured with {\it Chandra} (Naz\'e et al.\,\cite{NRM}).

Measuring the stellar RVs was, once more, quite difficult. The reasons for this were (1) the contamination of most, sufficiently intense, stellar lines (H\,{\sc i} and He\,{\sc i}) by rather strong nebular emission lines, and (2) the moderate S/N of our data that hampers the detection of very faint spectral lines. At first sight, only the signature of one star is seen in the spectra. The RVs of this main component vary very little during our observations (which cover {\it a priori} both quadrature phases of the orbit). This is very puzzling and prompted us to re-inspect the morphology of several lines in more detail. 

\begin{figure}[htb]
\resizebox{8.5cm}{!}{\includegraphics{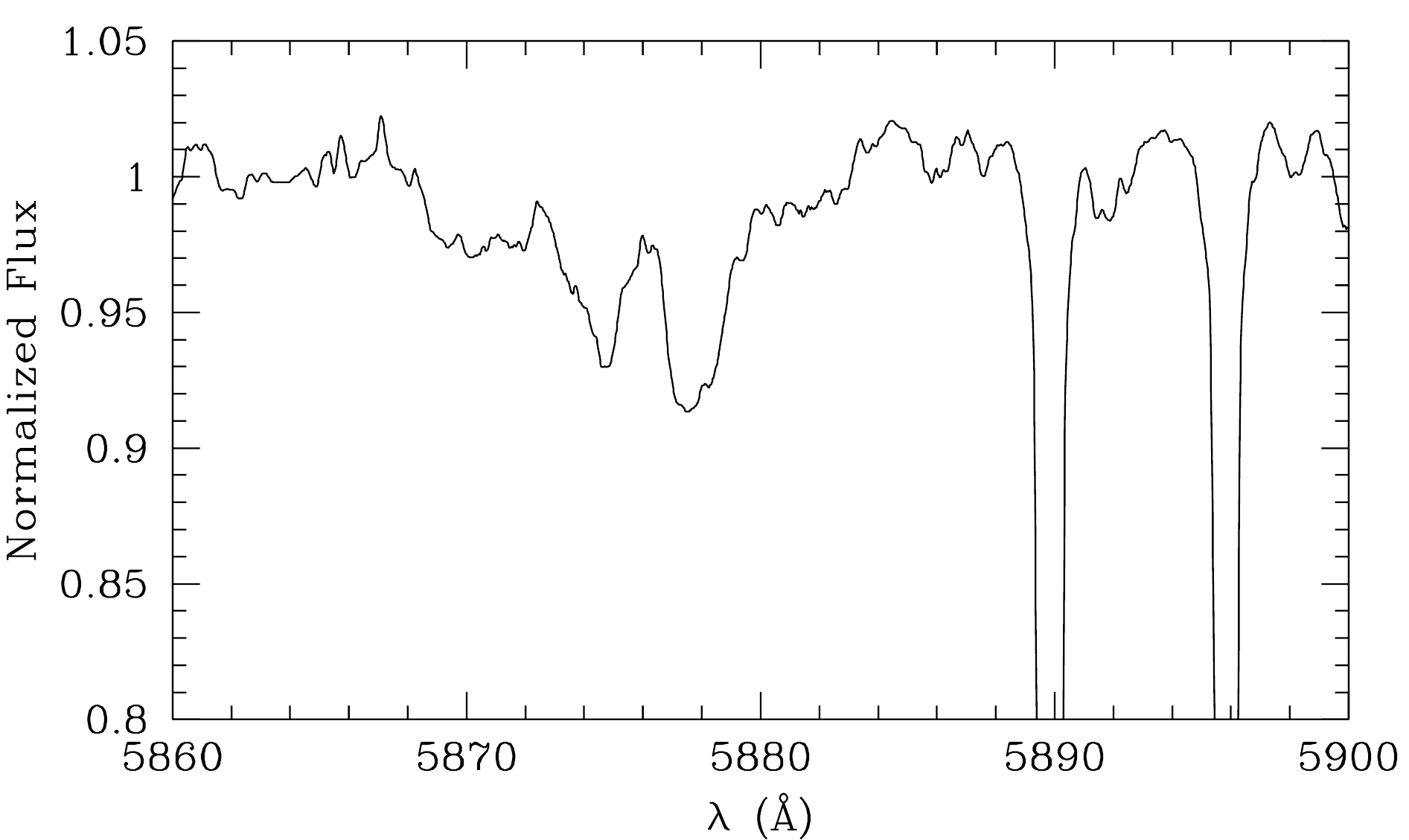}}
\caption{The region around the He\,{\sc i} $\lambda$\,5876 line in the spectrum of MSP\,44 observed on HJD\,2454539.622. The emission peak near 5876\,\AA\ is the nebular He\,{\sc i} line. Note the complex morphology of the line, especially in the blue and red wing.\label{specMSP44}}
\end{figure}

The most intriguing line turned out to be He\,{\sc i} $\lambda$\,5876. The blue and red wings of this line suggest the presence of two additional absorption features in most of our data (see Fig.\,\ref{specMSP44}). There are no known DIBs at these wavelengths (Herbig \cite{Herbig})\footnote{An up-to-date catalog of DIBs can be found at {\tt http://leonid.arc.nasa.gov/DIBcatalog.html}.} and the blue wing of the line at least should not be affected by any strong telluric absorption lines either (Curcio, Drummeter \& Knestrick \cite{telluric}). 

Since it is possible that MSP\,44 might be a triple system, we thus performed a multiple Gaussian fit of this line with up to four components (three stellar absorption components and one nebular emission component). The nebular component was consistently fitted with an RV of $(29.4 \pm 3.1)$\,km\,s$^{-1}$. The main stellar component was in turn found to display RVs between $17$ and $46$\,km\,s$^{-1}$. The corresponding RVs are listed in Table\,\ref{tablemsp44}. Although the RVs of the blue- and red-shifted absorption features must be interpreted with extreme caution, it is immediately clear that if MSP\,44 were a triple system, then the mass ratio of the binary responsible for the blue and red absorption components would be close to unity. This in turn would be at odds with the light curve, which implies that there is a rather large difference in surface brightness between the components of the eclipsing binary. The most likely interpretation of this line morphology therefore seems that it represents the very broad absorption line of only one star, but either deformed by pulsations such as in $\beta$\,Cep-type variables (e.g.\ Telting et al.\ \cite{Telting}) or partially filled in by some circumstellar emission or chromospheric emission from a low-mass companion heated by the irradiation from the B-star. In the case of the former interpretation, it must be stressed that our photometry does not reveal the signature of typical $\beta$\,Cep variations (see Paper I). As to the circumstellar emission interpretation, we note that neither H$\beta$ nor H$\alpha$ have emission above the continuum, but both lines have a somewhat similar morphology to that of He\,{\sc i} $\lambda$\,5876, in the sense that they exhibit a rather narrow core and more extended wings (especially the red wing).
   
In the absence of an unambiguous interpretation of the spectral morphology of MSP\,44, we choose to use the RVs of the He\,{\sc i} $\lambda$\,4921 line, which is less complex than He\,{\sc i} $\lambda$\,5876, to derive some constraints on the orbital motion of the main component. Assuming that the RVs of this line indeed reflect orbital motion, a circular SB1 orbital solution yields $T_0 = 2454531.774 \pm 0.285$, $\gamma_1 = (42 \pm 9)$\,km\,s$^{-1}$, $K_1 = (34 \pm 16)$\,km\,s$^{-1}$, and $a_1\,\sin{i} = 3.5\,R_{\odot}$. This yields a very low mass-function of $f(m) = \frac{m_2^3\,\sin^3{i}}{(m_1 + m_2)^2} = (0.021 \pm 0.030)\,M_{\odot}$. Here, $T_0$ stands for the time of primary minimum. Compared to the photometric ephemerides, there is a shift in phase by $0.116$ or 1.116. Since the number of orbital cycles between our photometric and spectroscopic campaigns amounts to 224, this phase shift implies an error in the orbital period of either 0.003 or 0.026\,days, well within or consistent with the estimated error in the orbital period as inferred from the photometric data (0.029\,days, see above). 

Since the system is eclipsing, we can to first order approximate the $\sin^3{i}$ term in the mass function to be unity. Assuming a mass of 10 -- 12\,$M_{\odot}$ for the B1 primary, we find that the secondary should have a mass of about 1.4 -- 1.6\,$M_{\odot}$, which implies that the companion might be an early F-type star. But would this result be reasonable? To answer this question, we return to the photometric light curve of the system. We first recall that the latter displays variations outside the eclipses that are quite large. In Paper I, we attributed them to spots, but our current knowledge of the system allows us to be a bit more specific. If we are indeed dealing with an F star closely orbiting around a B1\,V primary, it seems unavoidable that there will be a strong heating effect of part of the secondary's surface by the primary's radiation. Test calculations carried out with the {\sc nightfall} code indicate that the light curve of the system can {\it a priori} be explained by assuming that the hemisphere of the secondary facing the primary is significantly hotter than the rear side. However, the solution of the light curve is certainly not unique, because we ignore the actual mass ratio of the system and this parameter cannot be well-constrained from the photometric data alone. As an illustration, we show below the best-fit solution obtained for $m_1/m_2 = 8$.  

\begin{figure}[htb]
\resizebox{8.5cm}{!}{\includegraphics{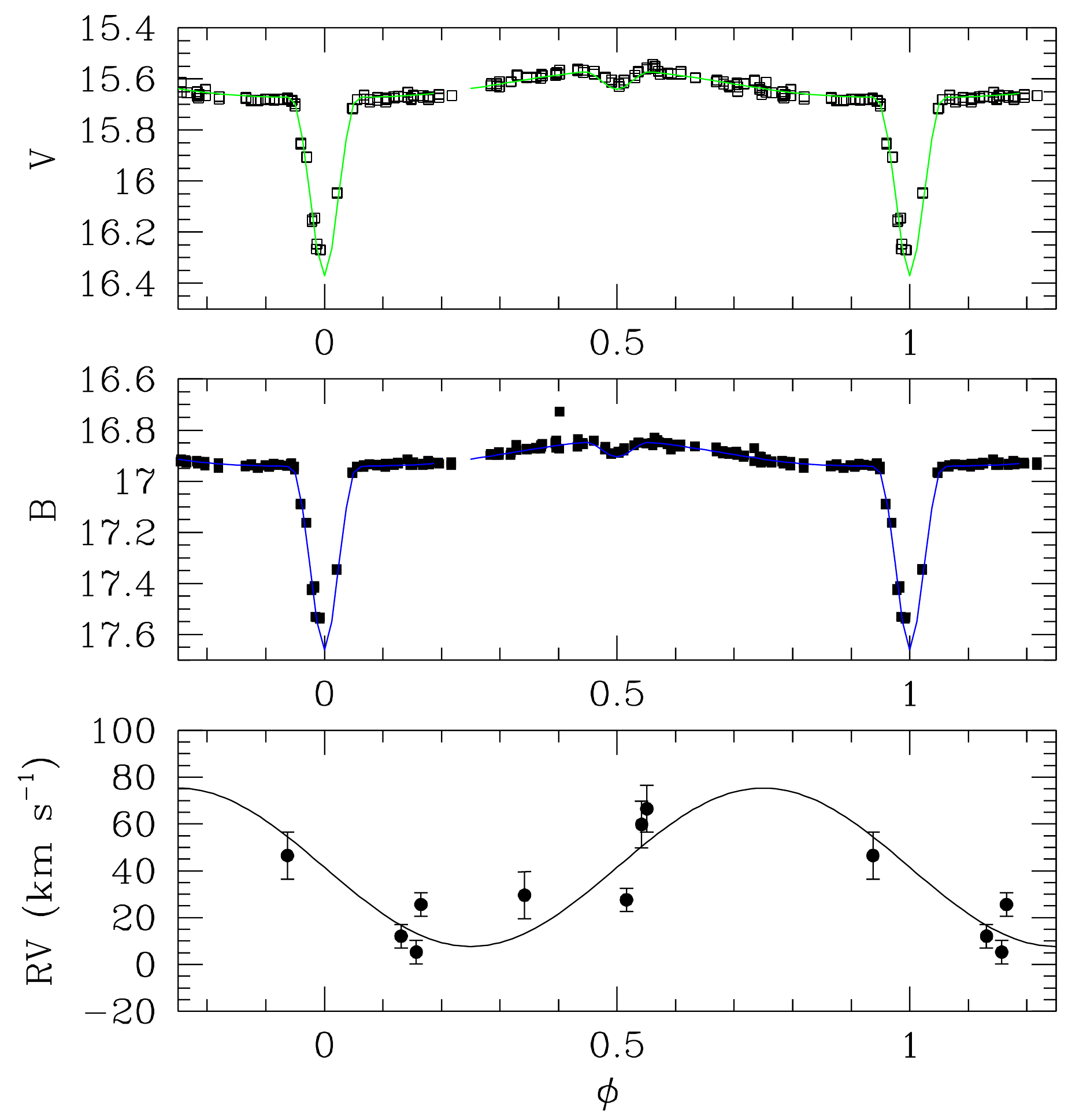}}
\caption{Combined solution of the light curve and the SB1 RV curve of MSP\,44. The parameters of the light curve model are $m_2/m_1 = 0.125$, $i = 80.6^{\circ}$, primary and secondary radii of 5.2 and 5.6\,$R_{\odot}$ respectively, and mean surface temperatures of $T_1 = 29100$\,K (fixed) and $T_2 = 6440$\,K. The secondary star features a hot spot facing the primary with a radius of $45^{\circ}$ and a temperature enhanced by a factor 1.76. \label{MSP44sol}}
\end{figure}

We note in addition that the radius of the secondary inferred from the light curve solution (5.6\,$R_{\odot}$ in the case of the best-fit solution shown below) is actually much larger than the expected radius of an early-F main-sequence star, but closer to what one would expect for a giant. Given the youth of the Westerlund\,2 cluster, the most straightforward explanation is that the secondary star of MSP\,44 is not a genuine giant, but rather a pre-main sequence (PMS) object. A PMS nature of the secondary would also provide a very natural explanation of the rather bright and hard X-ray emission of this system. 

In summary, our analysis of the spectroscopic and photometric data of MSP\,44 leaves a number of open questions about the exact nature of the primary component. The most likely scenario seems to be that this system consists of a B1 main-sequence star and a low-mass PMS companion. The physical parameters of the system are admittedly too uncertain to attempt a self-consistent determination of the distance of Westerlund\,2, although we note that the parameters of the fit in Fig.\,\ref{MSP44sol} are consistent with a distance of 8.4\,kpc. If the B1\,V spectral type of the MSP\,44 primary were correct and the companion contributed little to the integrated light of the system, then the $V$ magnitude and $B-V$ colour outside eclipse would indeed be consistent with a distance near 8.4\,kpc.
 
\subsection{MSP\,223}
\begin{figure}[htb]
\resizebox{8.5cm}{!}{\includegraphics{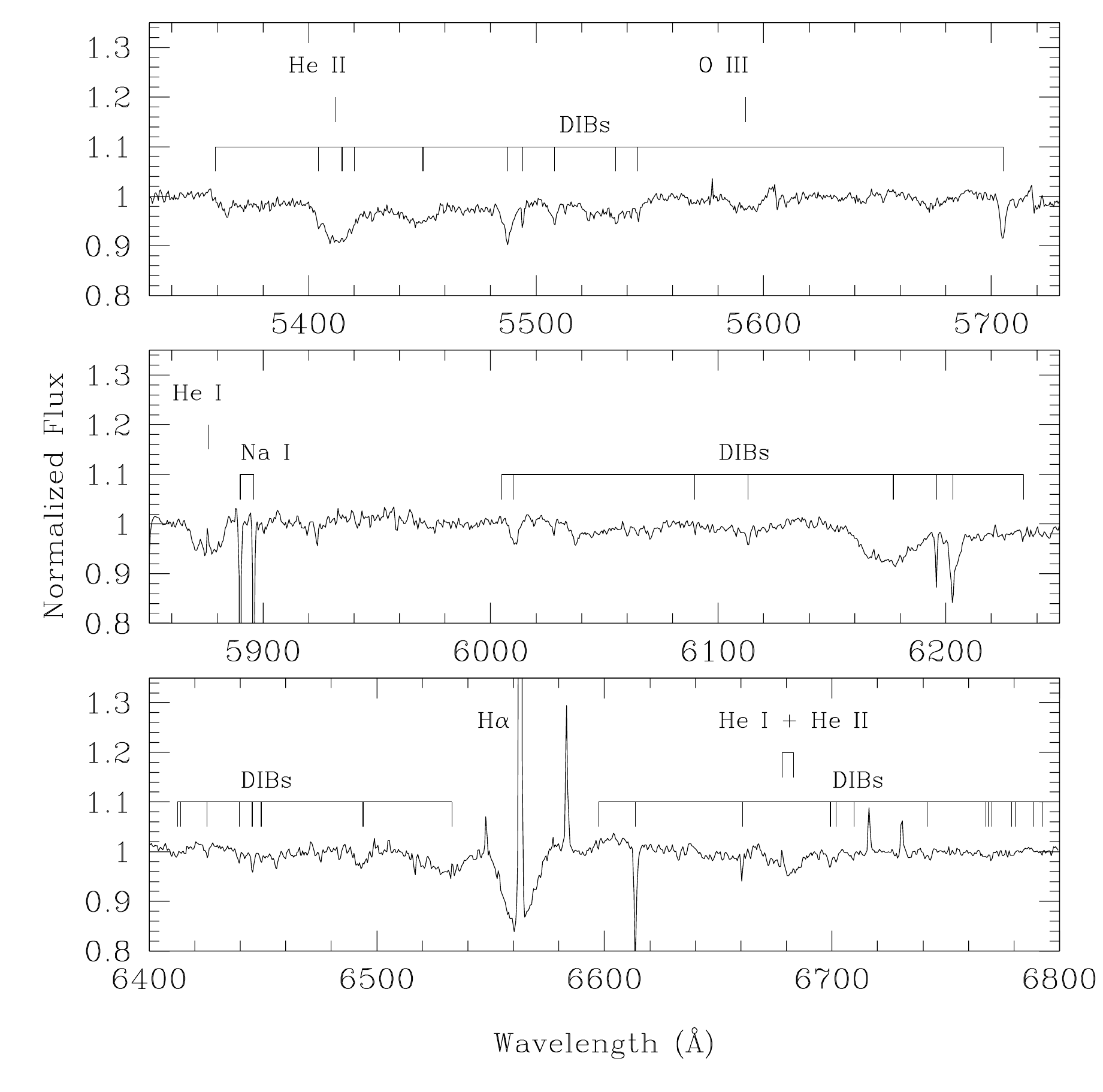}}
\caption{Mean spectrum of MSP\,223. Most of the features are due to interstellar lines or bands. The rather prominent He\,{\sc ii} $\lambda$\,5412 line indicates an O-star classification (see text for details).\label{spec223}}
\end{figure}

The yellow-red spectra of MSP\,223 display a rather strong He\,{\sc ii} $\lambda$\,5412 feature, as well as a weak (EW $\sim 0.4$\,\AA), but definitely detected O\,{\sc iii} $\lambda$\,5592 line (see Fig.\,\ref{spec223}). Whilst there are no known interstellar features affecting the O\,{\sc iii} line, there are three, rather weak, DIBs at $\lambda\lambda$\,5404.5, 5414.8, and 5420.2 that are blended with He\,{\sc ii} $\lambda$\,5412. To assess the real strength of the He\,{\sc ii} $\lambda$\,5412 feature, we first evaluated the ratio of the EWs for some nearby DIBs ($\lambda\lambda$\,5450.3, 5487.5) in the spectrum of MSP\,223 and the corresponding EWs in the spectrum of HD\,183143 quoted by Herbig (\cite{Herbig}). This ratio (1.33) was then used as a linear scaling factor to estimate the contamination of the He\,{\sc ii} $\lambda$\,5412 line by the three DIBs. The corrected EW of the He\,{\sc ii} line amounts to 0.9\,\AA. Therefore, He\,{\sc ii} $\lambda$\,5412 is of comparable strength to the He\,{\sc i} $\lambda$\,5876 line (EW = 0.85\,\AA), although the latter is slightly affected (probably less than 0.2\,\AA) by a nebular emission line. These properties suggest an O7-8 spectral type (Walborn \cite{NRW}) for the combined spectrum. That we are dealing with an O-type star is confirmed by the absence of He\,{\sc i} $\lambda$\,4921 and the fact that the He\,{\sc i} $\lambda$\,6678 line is clearly blended with He\,{\sc ii} $\lambda$\,6683. 

\begin{figure}[htb]
\resizebox{8.5cm}{!}{\includegraphics{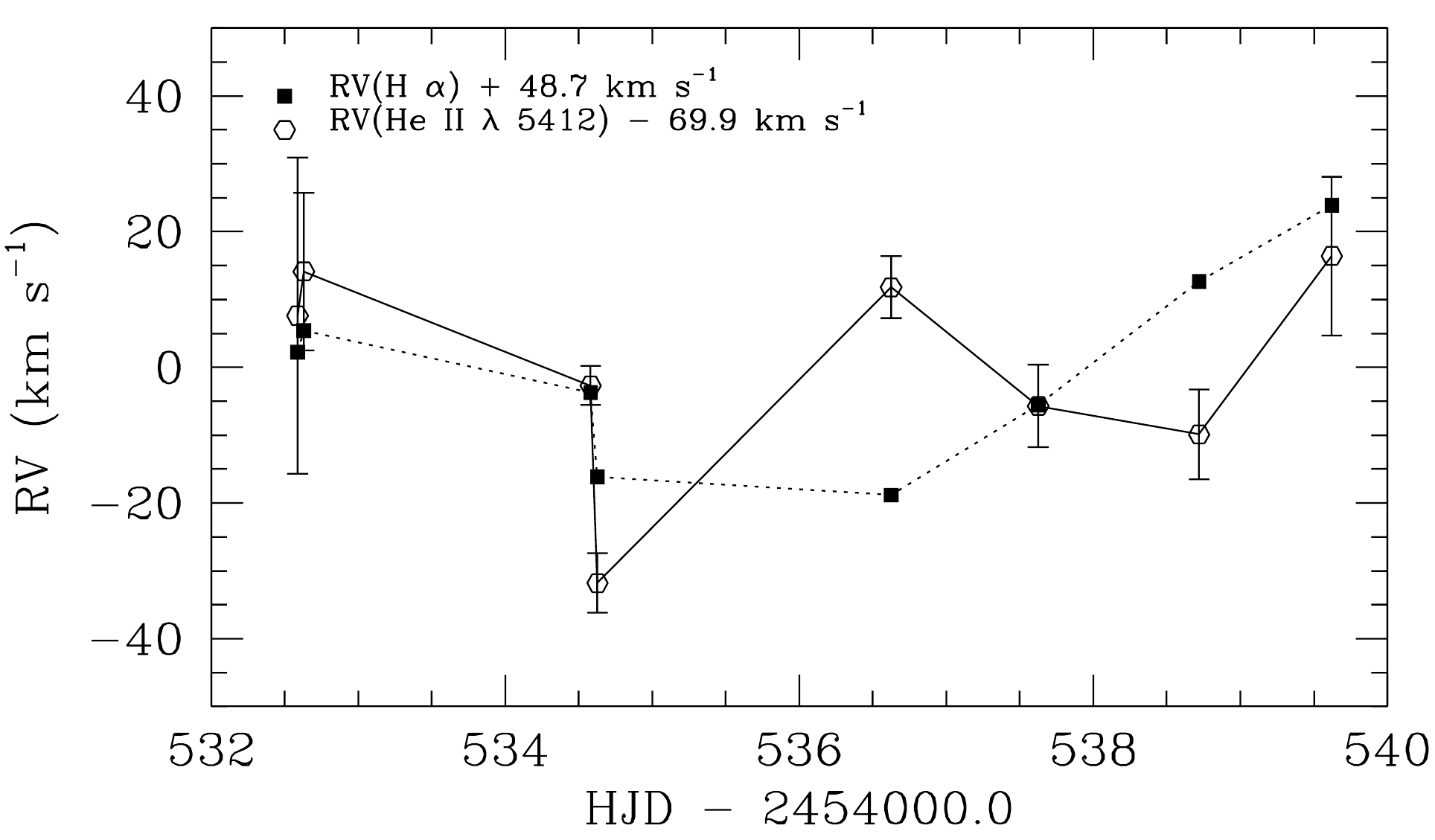}}
\caption{Radial velocity variations of MSP\,223 during our observations. The error bars in the H$\alpha$ RVs yield the formal error in the position of the absorption line in the two-Gaussian fit accounting for the nebular emission line.\label{RVs223}}
\end{figure}

We measured the RVs of several stellar lines. The most robust results were obtained for He\,{\sc ii} $\lambda$\,5412 and H$\alpha$, although we caution that the former is affected by blends with DIBs and for the latter, we had to perform a two Gaussian fit to account for the contamination by the nebular H$\alpha$ emission line\footnote{In these fits, the nebular line has a very stable RV of $1.1 \pm 0.2$\,km\,s$^{-1}$.}. The results are listed in Table\,\ref{tablemsp223} and plotted in Fig.\,\ref{RVs223}, after subtracting the mean RV individually for each line. No obvious trend is apparent in this figure. The rather large shift in RV between the He\,{\sc ii} $\lambda$\,5412 and H$\alpha$ line is most likely due to wind emission affecting the latter line. Similar shifts in the apparent systemic velocity of different lines are commonly found in binary systems where at least one component features a strong stellar wind (e.g.\ Rauw et al.\ \cite{hde228766}).

\begin{table}
\caption{Radial velocities of MSP\,223.\label{tablemsp223}}
\begin{center}
\begin{tabular}{c c c c}
\hline
HJD-2454000 & RV(He\,{\sc ii} $\lambda$\,5412) & RV(H$\alpha$) \\
            &  (km\,s$^{-1}$)                  & (km\,s$^{-1}$)\\
\hline
532.587 & 77.5 & $-46.4$\\
532.631 & 84.0 & $-43.2$\\ 
534.581 & 67.2 & $-52.4$\\
534.628 & 38.1 & $-64.8$\\
536.624 & 81.7 & $-67.5$\\
537.626 & 64.2 & $-54.2$\\
538.719 & 60.0 & $-36.0$\\
539.622 & 86.3 & $-24.8$\\
\hline
\end{tabular}
\end{center}
\end{table}

Our previous photometric monitoring of the cluster indicated that MSP\,223 is probably an eclipsing binary (Paper I). The Fourier analysis yielded the highest peak for a period of 14.61\,days, with a rather large uncertainty of 1.2\,days. The light curve folded with this period yields only a primary eclipse, with no indication of a secondary eclipse, although we might have missed it, if its duration is sufficiently short to fit into a small gap in phase coverage around $\phi = 0.5$. In Paper I, we inferred an orbital period of 29.211\,days, which then indicates two rather similar eclipses. This would imply that the system consists of two rather similar stars, well within their Roche lobe and seen at a large inclination angle. However, we do not observe an obvious SB2 signature in our spectra of MSP\,223, which suggests instead that the companion of the O-star is significantly fainter and less massive. This would be consistent with the rather modest RV variations that we have measured (see Fig.\,\ref{RVs223}), which show a peak-to-peak amplitude of about 40\,km\,s$^{-1}$ over seven days (the most reliable RVs are probably those of H$\alpha$). Unfortunately, the considerable uncertainties in the orbital period prevent us from extrapolating the photometric ephemerides to the epoch of our spectroscopic observations and we cannot derive constraints on the orbital solution of this system.  

Although our knowledge about MSP\,223 is still rather poor, we can nevertheless attempt to derive its spectro-photometric distance. For this purpose, we combine the photometric information from Paper I ($V = 15.75$, $B-V = 1.50$) with the absolute magnitudes and intrinsic colours of O7-8\,V stars from Martins \& Plez (\cite{MP}). In this way, we obtain a distance modulus of 14.67 -- 14.96 corresponding to a distance of 8.6 -- 9.8\,kpc. If our spectral classification were too early, and we adopted instead an O9.5\,V type, we would still derive a distance of 7.2\,kpc. We stress that this number does not account for the contribution of a companion to the light of MSP\,223. Accounting for this would lead us to inferring an even larger distance.

\subsection{MSP\,171, 182, and 203 + 444}
\begin{figure}[htb]
\resizebox{8.5cm}{!}{\includegraphics{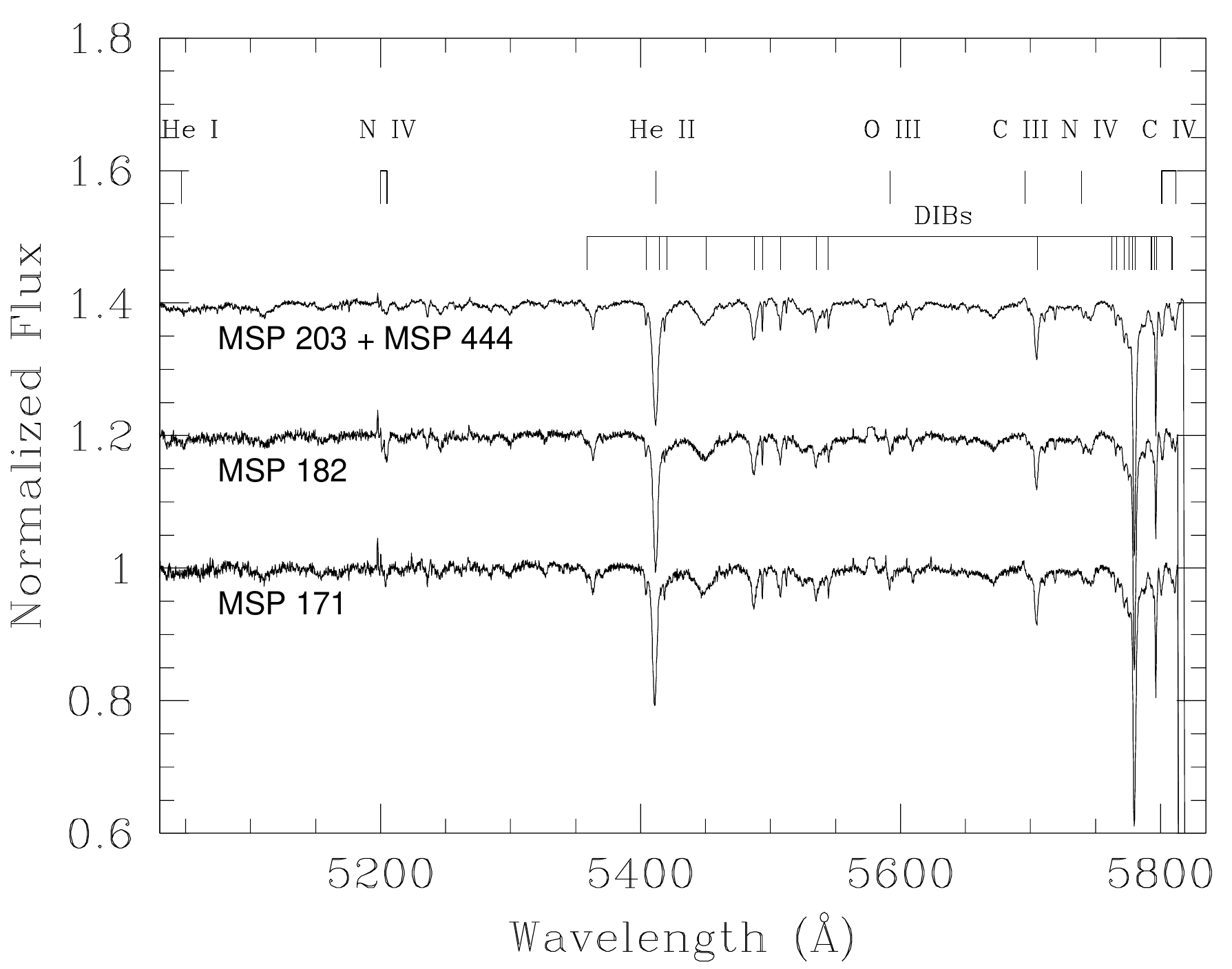}}
\caption{Mean GIRAFFE spectra of MSP\,171, 182, and 203 + 444.\label{ifu}}
\end{figure}

These three (actually four) objects were observed with the integral field spectroscopy units connected to the GIRAFFE spectrograph. Their mean spectra are shown in Fig.\,\ref{ifu}.

In Paper I, the combined blue spectrum of MSP\,203 + MSP\,444 was assigned a spectral type O6\,V-III. However, the GIRAFFE spectra of MSP\,203 + MSP\,444 did not reveal any trace of He\,{\sc i} $\lambda\lambda$\,5016, 5048. Therefore, by comparison with the yellow-green spectra of OB stars published by Kerton, Ballantyne \& Martin (\cite{Kerton}), the brightest star of the blend must have a spectral type earlier than O6\,V and possibly even earlier than O5\,V. The mean spectrum of MSP\,203 + MSP\,444 is very similar to that of MSP\,171 and 182, which were classified as O4-5 and O4 respectively in Paper I. We thus conclude that the brighter component of MSP\,203 + MSP\,444\footnote{According to the photometry of Moffat et al.\ (\cite{MSP}), MPS\,203 is slightly brighter than MSP\,444. However, the HST data (see Fig.\,\ref{argus}) suggest the reverse situation.} is likely of spectral type O4-5\,V. The other component is likely responsible for the weak He\,{\sc i} $\lambda$\,4471 absorption that was seen in Paper I. We note that the latter absorption is significantly broader than the other stellar absorption lines in the blue (and yellow-green) spectra of MSP\,203 + MSP\,444. This supports our suggestion that this line is not associated with the same star as the stronger, higher ionization lines. 

We measured the RVs of the three stars by fitting Gaussian profiles to the He\,{\sc ii} $\lambda$\,5412, O\,{\sc iii} $\lambda$\,5592, and C\,{\sc iv} $\lambda\lambda$\,5801, 5812 lines with the caveat that the C\,{\sc iv} lines (especially C\,{\sc iv} $\lambda$\,5812) are heavily blended with DIBs. We also measured the RV dispersions of four DIBs ($\lambda\lambda$ 5362, 5487, 5705, and 5781, see Herbig \cite{Herbig}), which are of comparable strength to or slightly stronger than the stellar lines. 

For MSP\,171, we infer a $1 - \sigma$ RV dispersion of 6.9\,km\,s$^{-1}$ for the strongest stellar line (He\,{\sc ii} $\lambda$\,5412) and 3.5\,km\,s$^{-1}$ for the C\,{\sc iv} $\lambda$\,5812 line. For the same set of spectra, the RV dispersions of the DIBs are between 0.8 and 10.3\,km\,s$^{-1}$ with a typical value of about 5\,km\,s$^{-1}$.
 
In the case of MSP\,182, all stellar lines, except for C\,{\sc iv} $\lambda$\,5812, which is affected by the DIB at 5809\,\AA\ (Herbig \cite{Herbig}), display a $1 - \sigma$ RV dispersion in the range 3.0 -- 5.7\,km\,s$^{-1}$. For the interstellar bands, these dispersions are 1.1 -- 4.1\,km\,s$^{-1}$, except for the rather weak DIB at $\lambda$\,5362 which has a larger dispersion of 9.3\,km\,s$^{-1}$.

Finally, the strongest stellar lines of the MSP\,203 + MSP\,444 spectra have RV dispersions of 2.5 -- 4.1\,km\,s$^{-1}$, whilst the weak O\,{\sc iii} $\lambda$\,5592 feature has a slightly larger dispersion of 7.6\,km\,s$^{-1}$. In this case, the RV dispersions of all DIBs measured amount to 1.6 -- 3.3\,km\,s$^{-1}$. 

In conclusion, none of the stars observed with the IFUs feeding GIRAFFE is found to display significant RV variations on timescales of a week.

\subsection{Stars observed with the ARGUS unit \label{Sectargus}}
The field of view covered by the ARGUS unit harbours two stars with an MSP identifier (MSP\,167 and 183) as well as five anonymous objects (labeled A to E in Fig.\,\ref{argus}) that are bright enough to extract at least an averaged spectrum. In addition, one spaxel at the edge of the field contains photons from the wing of the PSF of star MSP\,199 which is right outside the nominal field of view. For the brightest and most isolated objects, the spectra were extracted over nine spaxels (MSP\,167, MSP\,183, object A), whilst the extraction region was reduced to a single spaxel for those stars (objects B, C, D, and E) that are either relatively close to a bright source or fall only partially inside the field of view of the ARGUS unit (MSP\,199). In this section we discuss the results obtained for the different stars in the ARGUS field. 

\begin{figure}[htb]
\resizebox{8.5cm}{!}{\includegraphics{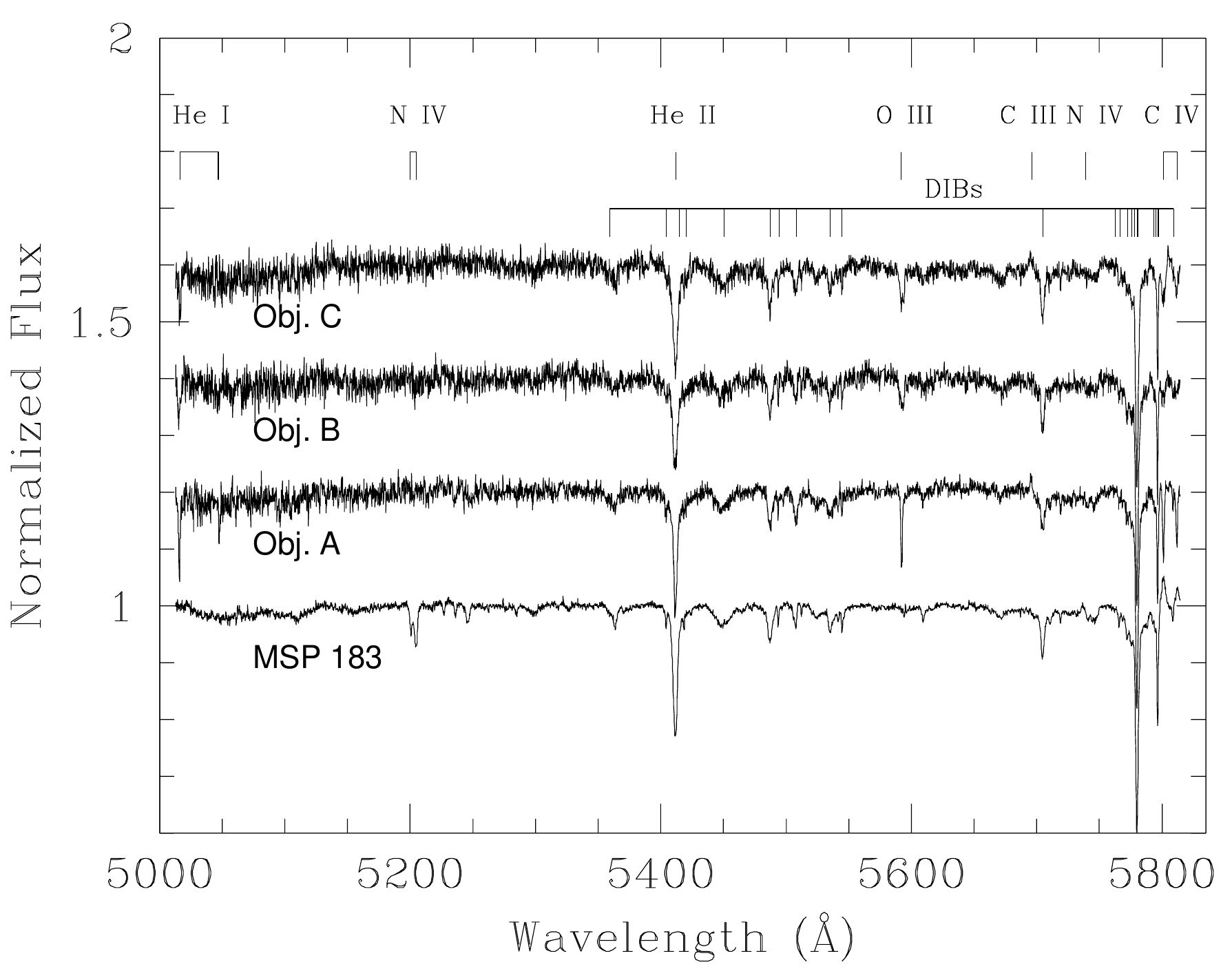}}
\caption{Mean GIRAFFE-ARGUS spectra of some of the targets discussed in Sect.\,\ref{Sectargus}.\label{argusspec}}
\end{figure}

In Paper I, MSP\,183 was classified as an O3\,V((f)) star. The lack of He\,{\sc i} $\lambda\lambda$\,5015, 5047 absorption, the strength of the He\,{\sc ii} $\lambda$\,5412 line 
%(EW = 0.84\,AA\ as derived from a simultaneous fit of the stellar line and the surrounding DIBs) 
and the presence of the N\,{\sc iv} $\lambda\lambda$\,5200 - 5205 blend in the GIRAFFE-ARGUS spectrum of MSP\,183 (see Fig.\,\ref{argusspec}) are in line with this classification (Kerton et al.\ \cite{Kerton}). However, the most remarkable feature of the spectrum of MSP\,183 is the presence of the C\,{\sc iv} $\lambda\lambda$\,5801, 5812 lines in weak, but definite, emission with equivalent widths of $-0.24$\,\AA\ and $-0.06$\,\AA, respectively. These lines were reported to be in emission in the spectrum of the O3\,If$^*$ supergiant Cyg\,OB2 \#7 (Walborn \& Howarth \cite{WH}). Our blue spectrum of MSP\,183 (see Paper I) rules out the possibility that MSP\,183 could be such an evolved supergiant. Therefore, it seems more likely that these C\,{\sc iv} emission lines are not selective properties of O3\,If$^*$ supergiants, but could actually be a general feature of the hottest O-type stars as suggested by Walborn (\cite{Walborn}).

The RVs of MSP\,183 were measured by fitting Gaussian profiles to the strongest stellar lines (He\,{\sc ii} $\lambda$\,5412, O\,{\sc iii} $\lambda$\,5592, C\,{\sc iv} $\lambda\lambda$\,5801, 5812; see Table\,\ref{RVMSP183}). In this way, we obtain a $1 - \sigma$ RV dispersion of 1.2\,km\,s$^{-1}$ for the He\,{\sc ii} $\lambda$\,5412 absorption and 7.3 and 3.3\,km\,s$^{-1}$ for the C\,{\sc iv} $\lambda$\,5801 and $\lambda$\,5812 emission lines respectively. The RV dispersion of the DIBs range between 0.9 and 7.5\,km\,s$^{-1}$ depending on the strength and shape of the DIB. We therefore conclude that this star does not display any significant RV variations during our observations. 

On the basis of a spectrum in the blue domain, we previously classified MSP\,167 as an O6\,III star (Paper I). The yellow spectral domain, however, is more consistent with an earlier spectral type. Indeed, our FLAMES-ARGUS spectra reveal no trace of either the He\,{\sc i} $\lambda\lambda$\,5015, 5047 or the O\,{\sc iii} $\lambda$\,5592 absorption lines. Furthermore, the C\,{\sc iv} $\lambda\lambda$\,5801, 5812 lines are clearly not present in absorption. If anything, C\,{\sc iv} $\lambda$\,5801 is in very weak emission (EW $\sim -0.10$\,\AA). These properties are at odds with an O6 spectral type because the  O\,{\sc iii} $\lambda$\,5592 and C\,{\sc iv} $\lambda\lambda$\,5801, 5812 absorption lines are expected to be strongest for the spectral types O6-O7 (Walborn \cite{NRW}). 

\begin{figure}[htb]
\resizebox{8.5cm}{!}{\includegraphics{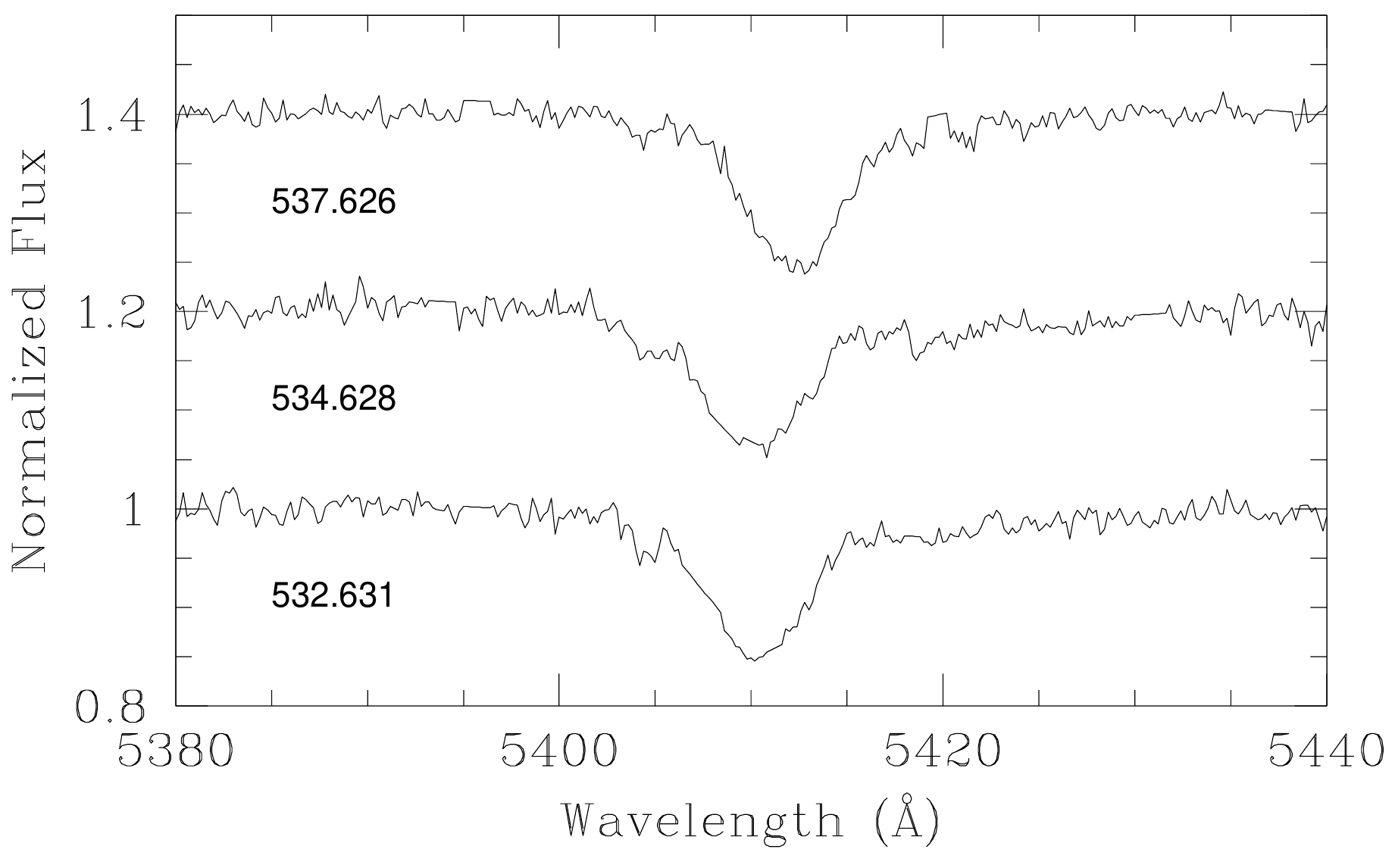}}
\caption{The He\,{\sc ii} $\lambda$\,5412 line in the spectrum of MSP\,167 at the three dates of our GIRAFFE-ARGUS observations. The shift between the second and the third observation is clearly seen.\label{msp167}}
\end{figure}

The dominant stellar line is He\,{\sc ii} $\lambda$\,5412. We measured its RVs by fitting Gaussian profiles (see Table\,\ref{RVMSP167}). Whilst there is only little difference between the RVs of the first two spectra, there is a clear shift of more than $100$\,km\,s$^{-1}$ between the second and third observation (see Fig.\,\ref{msp167}). This shift is highly significant since the RV dispersion of the DIBs range between 1.0 and 9.5\,km\,s$^{-1}$ with the lower limit on the dispersion holding for those DIBs which are of similar strength to the He\,{\sc ii} $\lambda$\,5412 line.  
\begin{table}[htb]
\caption{Radial velocities of the binary candidate MSP\,167.\label{RVMSP167}}
\begin{center}
\begin{tabular}{c c }
\hline
HJD-2454000        & \multicolumn{1}{c}{RV (He\,{\sc ii} $\lambda$\,5412)} \\
                   & (km\,s$^{-1}$) \\
\hline
532.631 & $-56.2$ \\
534.628 & $-64.5$ \\
537.626 &   49.0  \\
\hline
\end{tabular}
\end{center}
\end{table}

The above results suggest that MSP\,167 is actually a short-period (of the order of a week) spectroscopic binary consisting of an early O-type star (responsible for the strong He\,{\sc ii} line seen in our ARGUS data) and a companion of later spectral type (which produces the He\,{\sc i} absorption lines seen in the blue-band spectra). We note that this star displays a rather large X-ray luminosity of $8.9 \times 10^{32}$\,erg\,s$^{-1}$ with some hints of moderate ($2\,\sigma$) variability (see Naz\'e et al.\ \cite{NRM}). These properties, along with the RV variations reported above, suggest that MSP\,167 might be an interacting-wind binary system.\\

For MSP\,199 (O3-4\,V according to Paper I), the spectra were extracted from a single spaxel of the ARGUS unit. Therefore, the S/N of the individual spectra is rather low. Nevertheless, the combined spectrum agrees with the classification proposed in Paper I. The RVs of the He\,{\sc ii} $\lambda$\,5412 display a $1 - \sigma$ dispersion of 5.3\,km\,s$^{-1}$, whilst the DIBs at $\lambda$\,5781 and $\lambda$\,5796 have RV dispersions of 1.9 and 2.4\,km\,s$^{-1}$, respectively.\\ %Therefore, we conclude that this star does not display any significant RV variations during our observations. 

Objects A, B, C, D, and E are discussed here for the first time. Our mean spectrum of object A (10:24:02.8 $-$57:45:30.7, J2000) displays rather strong He\,{\sc ii} $\lambda$\,5412, O\,{\sc iii} $\lambda$\,5592, and C\,{\sc iv} $\lambda\lambda$\,5801, 5812 absorption lines along with fainter absorption lines of He\,{\sc i} $\lambda\lambda$\,5016, 5047 and a very weak C\,{\sc iii} $\lambda$\,5696 emission line (see Fig.\,\ref{argusspec}). These spectral features are indicative of an O8 spectral classification with an uncertainty of one subtype (Kerton et al.\ \cite{Kerton}, Walborn \cite{NRW}). The weakness of the C\,{\sc iii} emission most likely indicates a main-sequence luminosity class. Our photometric data (Paper I) indicate $V = 15.766 \pm 0.050$ and $B - V = 1.456$ for this star. Assuming $R_V = 3.1$ and adopting the intrinsic colours and absolute magnitudes of an O8\,V star (with an uncertainty of one spectral type) from Martins \& Plez (\cite{MP}), we infer a distance modulus $14.83 \pm 0.30$ corresponding to a spectrophotometric distance of ($9.2 \pm 1.3$)\,kpc. The RVs of the various stellar absorption lines yield $1 - \sigma$ dispersions in the range 1.1 to 6.5\,km\,s$^{-1}$ depending on the strength of the line considered (see Table\,\ref{RVobjA}). The lowest dispersions are found for the He\,{\sc ii} $\lambda$\,5412 and C\,{\sc iv} $\lambda$\,5801 lines. The DIBs at $\lambda$\,5781 and $\lambda$\,5796 have RV dispersions of 4.1 and 1.3\,km\,s$^{-1}$, respectively.\\

For star B (10:24:02.6 $-$57:45:31.2, J2000), the lines of He\,{\sc i} and O\,{\sc iii} are weak, but clearly present, whilst the C\,{\sc iv} absorption lines are also rather weak (see Fig.\,\ref{argusspec}). The strength of the He\,{\sc ii} $\lambda$\,5412 line implies a spectral type around O6-7. The RVs of the latter line indicate a $1 - \sigma$ dispersion of 3.8\,km\,s$^{-1}$, which is slightly larger than the one of the DIB at $\lambda$\,5781 (1.9\,km\,s$^{-1}$).\\ 

Star C (10:24:02.5 $-$57:45:32.1, J2000) displays a weak He\,{\sc i} $\lambda$\,5016 line, as well as shallow and broad O\,{\sc iii} and C\,{\sc iv} absorption lines (Fig.\,\ref{argusspec}), and C\,{\sc iii} $\lambda$\,5696 is seen in very weak emission. The spectral type is likely about O7\,V. The RV dispersion of the He\,{\sc ii} $\lambda$\,5412 line amounts to 2.2\,km\,s$^{-1}$, very much comparable to that of the DIBs at $\lambda$\,5781 and $\lambda$\,5797 (2.3 and 1.6\,km\,s$^{-1}$, respectively).\\ 

For stars D and E, the S/N of the spectra is rather low, preventing us from measuring the RVs of the stars from the individual spectra. We instead combined the three observations of each star into a mean spectrum with the goal of performing a spectral classification. Object D (10:24:02.2 $-57$:45:32.0, J2000) has a stronger He\,{\sc i} $\lambda$\,5016 than He\,{\sc ii} $\lambda$\,5412 line. We accordingly classify this star as O9.5 (Kerton et al.\ \cite{Kerton}). Star E (10:24:02.6 $-$57:45:32.9, J2000) must be of earlier spectral type, since there is no trace of He\,{\sc i} $\lambda$\,5016, whilst He\,{\sc ii} $\lambda$\,5412 is the strongest stellar line in the spectrum. A weak O\,{\sc iii} $\lambda$\,5592 line is seen, but no C\,{\sc iv} absorption could be detected. We suggest a spectral type near O6-7.\\ 

We inspected the {\it Chandra} image of the cluster's core (see Naz\'e et al.\ \cite{NRM}) at the positions of objects A -- E. Star E corresponds to the southwest lobe of an X-ray source detected between MSP\,183 and MSP\,199 using the wavedetect algorithm (Naze et al.\ \cite{NRM}). Whilst there is no obvious X-ray emission from stars A and C, there might be some X-ray emission associated with stars B and D. However, the latter two objects lie in the wings of the X-ray source associated with MSP\,183, and their status as X-ray emitters can therefore not be ascertained with 100\% confidence. The lack of X-ray detections for stars A and C is somewhat surprising given their spectral types inferred above. One possibility could be a larger interstellar column density, thus larger absorption of the X-ray emission than for other stars in Westerlund\,2 of similar spectral type. However, this explanation needs to be confirmed using additional high angular resolution optical and X-ray observations. 

\begin{table*}[htb]
\begin{center}
\caption{Results of our Monte Carlo simulations of the detection of massive binaries with the sampling of the GIRAFFE-IFU campaign.\label{MC}}
\begin{tabular}{c c c c c c}
\hline
$m_1$ ($M_{\odot}$) & I & II & III & IV & V \\
                    & $P \leq 5$ & $5 < P \leq 10$ & $10 < P \leq 20$ & $20 < P \leq 50$ & $50 < P \leq 100$ \\
& (\%) & (\%) & (\%) & (\%) & (\%) \\
\hline
20 & 0.7 & 0.7 & 1.8 & 19.0 & 56.2 \\
30 & 0.6 & 0.6 & 1.5 & 15.8 & 50.8 \\
40 & 0.5 & 0.5 & 1.2 & 13.8 & 46.6 \\
50 & 0.4 & 0.4 & 1.0 & 12.4 & 43.6 \\
\hline
\end{tabular}
\end{center}
\tablefoot{For a given range of orbital periods (expressed in days) and a given value of the primary mass, the table yields the fraction of systems (in \% of the population of systems within this interval of periods) that would display $\Delta$\,RV = $\max{RV(\phi_i)} - \min{RV(\phi_i)} < 20$\,km\,s$^{-1}$, hence escape detection. The total numbers of simulated systems in each of the period intervals are 34198, 25802, 4786, 6315, and 4772 for the intervals I, II, III, IV, and V, respectively.}
\begin{center}
\caption{Same as Table\,\ref{MC}, but for the sampling of the GIRAFFE-ARGUS data set. \label{MC2}}
\begin{tabular}{c c c c c c}
\hline
$m_1$ ($M_{\odot}$) & I & II & III & IV & V \\
                    & $P \leq 5$ & $5 < P \leq 10$ & $10 < P \leq 20$ & $20 < P \leq 50$ & $50 < P \leq 100$ \\
& (\%) & (\%) & (\%) & (\%) & (\%) \\
\hline
20 & 3.4 & 1.5 & 4.5 & 29.8 & 70.8 \\
30 & 2.7 & 1.2 & 3.4 & 26.2 & 64.8 \\
40 & 2.4 & 1.0 & 2.8 & 23.7 & 60.7 \\
50 & 2.1 & 0.9 & 2.3 & 21.7 & 57.6 \\
\hline
\end{tabular}
\end{center}
\end{table*}

\section{Discussion \label{discussion}}
The various eclipsing binaries that we have studied in this paper yield distance estimates in the range 6.5 to larger than 9\,kpc. Whilst the uncertainty in these results is difficult to evaluate, they are in very good agreement with the spectrophotometric distance of $(8.0 \pm 1.4)$\,kpc inferred in Paper I and clearly consistent with a cluster distance well beyond the 2.8\,kpc proposed by Ascenso et al.\ (\cite{Ascenso}). We thus conclude that all available observations of the early-type stellar population of Westerlund\,2 are most consistent with a distance near 8.0\,kpc in agreement with the distance inferred for WR\,20a. This result casts serious doubts on a possible association between the $\gamma$-ray pulsar PSR\,J1023.0$-$5746 (Ackermann et al.\ \cite{Ackermann}) and the Westerlund\,2 cluster. If the distance of the pulsar were confirmed at 2.4\,kpc, it would then appear more likely that the latter belongs to a foreground stellar population unrelated to the massive stars of Westerlund\,2, but possibly related to the PMS stars discussed by Ascenso et al.\ (\cite{Ascenso}).\\

With respect to the multiplicity of early-type stars in Westerlund\,2, we stress that among the eleven O-type stars and one WNha star monitored during our campaign, which were not previously known to be binary systems, only one star (MSP\,167) was found to probably be a short-period spectroscopic binary system. The other targets did not display RV variations well above the uncertainties estimated from the RV dispersion of the DIBs. To assess the significance of our results, we performed Monte Carlo simulations (Hammersley \& Handscomb \cite{HH}) of the RVs of a synthetic population of massive binaries. The parent distribution of the binary parameters of the population of massive binaries is unfortunately unknown, but we can make a series of reasonable assumptions. Here, we have followed an approach similar to the one of Sana, Gosset \& Evans (\cite{SGE}). Following the latter authors, we adopt a bi-uniform period distribution in $\log{P(days)}$ with 60\% of the systems in the range $0.3 \leq \log{P} \leq 1.0$ and the remainder in the range $1.0 < \log{P} \leq 3.5$. The eccentricity was assumed to be uniformly distributed between 0.0 and 0.9, with all systems with orbital periods shorter than four days assumed to have circular orbits. The longitudes of periastron and true anomaly of the first observation were taken to be uniformly distributed in the $[0,2\,\pi]$ interval and we adopted a uniform distribution of the mass ratio $q = m_2/m_1$ between 0.1 and 1.0. The lower limit to the range of mass ratio was chosen because, in practice, it would be very difficult to observe the reflex motion of an O-star for such a low-mass companion. Finally, the orbital inclination was taken to be uniform in $\cos{i}$ between $-1.0$ and $1.0$. 

\begin{figure}[htb]
\resizebox{8.5cm}{!}{\includegraphics{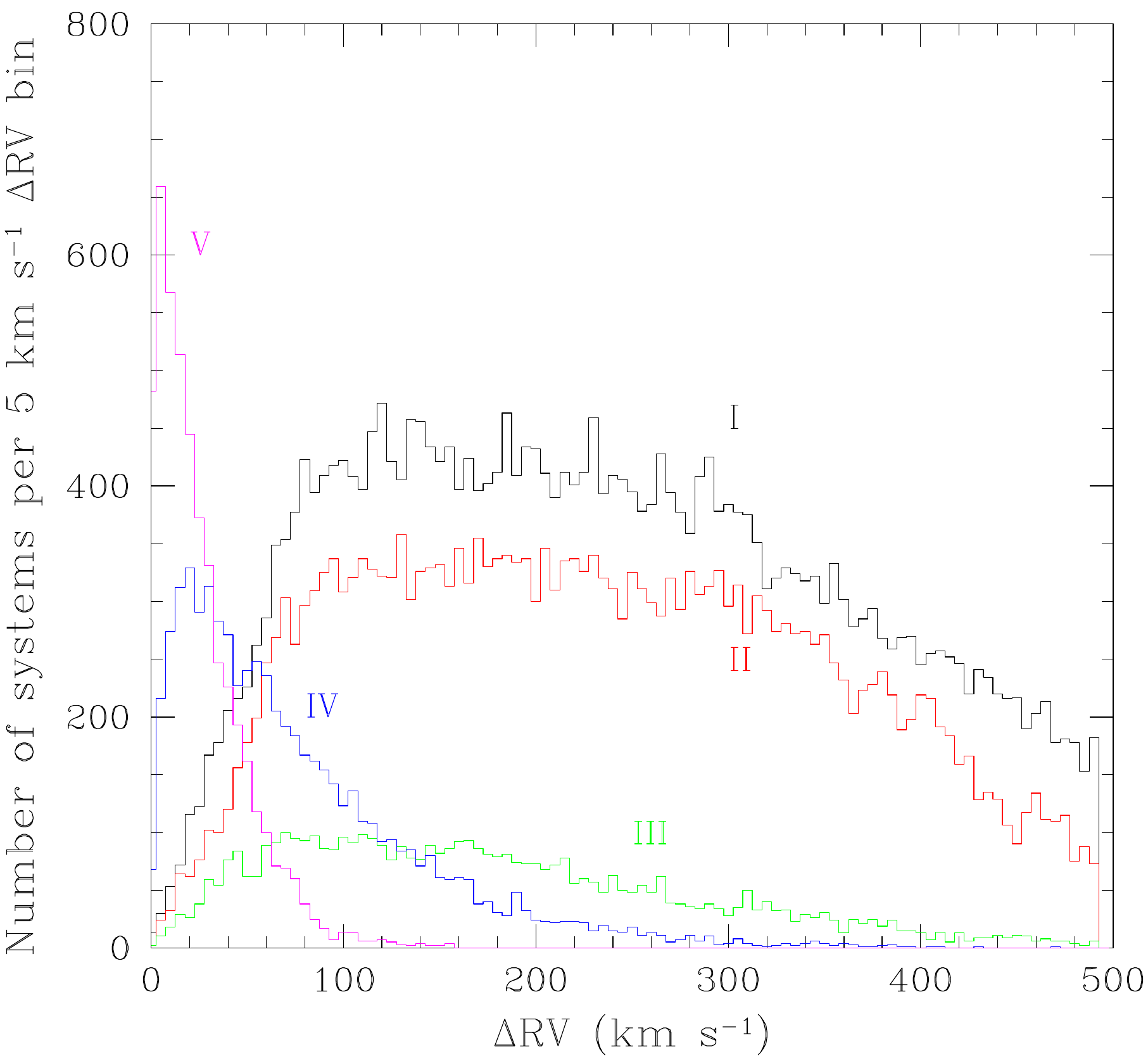}}
\caption{Distribution of $\Delta$\,RV = $\max{RV(\phi_i)} - \min{RV(\phi_i)}$ for a population of massive binaries subdivided into various intervals of orbital period (labelled I, II, III, IV, and V, see Table\,\ref{MC}). In this simulation, $m_1$ was assumed to be equal to 40\,$M_{\odot}$.\label{histo}}
\end{figure}

For four values of the primary mass (20, 30, 40, and 50\,$M_{\odot}$, spanning roughly the range of spectral types for which some observational information on multiplicity is available) we simulated a population of 100000 binary systems. For each system, we evaluated the maximum RV difference $\Delta$\,RV = $\max{RV(\phi_i)} - \min{RV(\phi_i)}$ that would be measured on this system adopting the same temporal sampling as used for our five GIRAFFE-IFU observations\footnote{Our UVES campaign includes eight pointings. However, the three additional pointings only yield a marginal improvement in the detection efficiency compared to the results discussed here.}. The results are summarized in Table\,\ref{MC} and illustrated in Fig.\,\ref{histo}. If we assumed that a velocity difference of 20\,km\,s$^{-1}$ were required to claim the detection of a binary system\footnote{None of the presumably single O-star objects discussed in this paper displays a velocity excursion of more than 20\,km\,s$^{-1}$ in the most stable photospheric lines.}, we would find that fewer than 1.5\% of the systems containing an O4-6\,V primary star (i.e.\ with a mass higher than 30\,$M_{\odot}$), and having an orbital period of shorter than 20\,days, would escape detection. Only the lowest inclination systems would indeed remain undetected. For longer orbital periods, the fraction of systems that would be missed by our GIRAFFE-IFU campaign steeply increases: for systems with orbital periods between 50 and 100\,days, about half of the binaries would escape detection. However, owing to the assumed preponderance of short orbital-period systems, the total fraction of systems with a period up to 100\,days that we would miss, would still remain quite low (4 - 5\%). 

The same approach was applied to the sampling of the three GIRAFFE-ARGUS observations (see Table\,\ref{MC2}). As could be expected, this sampling is less efficient for the detection of binary systems. Nonetheless, for primary masses in the range 20 - 50\,$M_{\odot}$ (corresponding roughly to O4-8\,V spectral types), our simulations indicate that a binary system with a period shorter than 100\,days has less than 10\% a probability of escaping detection because $\Delta$\,RV $< 20$\,km\,s$^{-1}$. 

Therefore, we can safely conclude that the probability that either of the stars MSP\,18, 171, 182, 183, 199, or 203 or stars A, B or C is a short-period binary system must be very low. If these stars are indeed binaries, they must have either a very low orbital inclination, a very long orbital period, or both. The situation is a bit less clear for WR\,20b for which we do observe a maximum RV difference above our threshold. However, as stated above, this result must be considered with caution, since all the strong emission lines display line-profile variability that could easily be responsible for the observed shifts in RV, whilst the weaker emission lines are found to be more constant, in terms of both morphology and RV. 

The lack of a clear binary signature for WR\,20b and MSP\,18 is especially surprising given their high X-ray luminosities (see Naz\'e et al.\ \cite{NRM}). These stars have $\log{L_{\rm X}/L_{\rm bol}} = -6.16$ and $-5.92$ for WR\,20b and MSP\,18 respectively, which are significantly larger than for typical single O-type stars and are as large as, or even larger than the value of the interacting wind system WR\,20a ($\log{L_{\rm X}/L_{\rm bol}} = -6.15$, Naz\'e et al.\ \cite{NRM}). Since colliding winds have the potential to generate copious amounts of X-rays (e.g.\ Stevens, Blondin \& Pollock \cite{SBP}), high X-ray luminosities have traditionally been associated with interacting winds in a binary system. However, recent investigations have revealed that interacting wind binaries are not always X-ray overluminous (e.g.\ Sana et al.\ \cite{Lx}, Naz\'e \cite{LxLbol}), at least as far as O-star binaries are concerned, and that alternative mechanisms such as magnetically confined winds can also produce high X-ray luminosities (e.g.\ Babel \& Montmerle \cite{BM}). This implies that colliding winds, and by extension binarity, are neither a necessary nor a sufficient condition for high X-ray luminosities. Nonetheless, the high X-ray luminosities of MSP\,18 and WR\,20b make these objects very interesting targets that deserve additional monitoring, in terms of both optical spectroscopy, to search for long-term RV variations, and spectropolarimetry, to search for evidence of magnetic fields or non-spherical winds.  

If these stars were indeed single, then their bolometric luminosities would suggest masses in the range 50 -- 70\,$M_{\odot}$, provided that the main-sequence mass-luminosity relation holds for all these objects. It has been argued that the most massive stars might actually display an O\,If$^*$ supergiant or hydrogen-rich WN type spectrum rather than a normal O-type spectrum (Crowther et al.\ \cite{Crowther}). While it is true that the highest ever dynamical masses have been measured for WNLha stars so far (Rauw et al.\ \cite{Rauw04}, Niemela et al.\ \cite{Niemela}, Schnurr et al.\ \cite{Schnurr08,Schnurr09}), it is puzzling that the WN6ha star WR\,20a and the O5\,V-III stars MSP\,18 and MSP\,203 have very similar bolometric luminosities and hence masses. In this context, it is worth recalling the result of De Becker et al.\ (\cite{IC1805}), and Taylor et al. (\cite{Taylor}). De Becker et al.\ (\cite{IC1805}) derived a minimum dynamical mass of $(150 \pm 50)\,M_{\odot}$ for the O5.5\,III(f) primary of HD\,15558, whilst Taylor et al.\ (\cite{Taylor}) inferred minimum dynamical masses of $M_1\,\sin^3{i} = (78 \pm 8)$ and $M_2\,\sin^3{i} = (66 \pm 7)\,M_{\odot}$ for the O6.5\,Iafc + O6\,Iaf binary system R\,139. After all, these results suggest that some apparently `normal' O-type stars could actually span a range of masses quite similar to that of WNLha stars.\\  

Although we have monitored only a limited sample of O-type stars in Westerlund\,2, our results suggest that unlike some other very young open clusters, such as NGC\,6231 (Sana et al.\,\cite{Sana}) or Trumpler\,16 (Rauw et al.\,\cite{Tr16}), Westerlund\,2 is most probably not very rich in short period early-type binaries. With only five binaries or binary candidates (including WR\,20a and the three eclipsing OB systems, two of which host B-type primaries) out of the fifteen stars investigated, our results are more reminiscent of the situation encountered in IC\,1805 (De Becker et al.\,\cite{IC1805}, Hillwig et al.\, \cite{Hillwig}) and NGC\,2244 (Mahy et al.\,\cite{Mahy}). If we restrict ourselves to O-type stars, the fraction of confirmed or candidate binaries drops to about 15\% (2 out of 13). Given the higher stellar density in the core of NGC\,6231 than in both NGC\,2244 and IC\,1805, Mahy et al.\ (\cite{Mahy}) tentatively suggested that the number of short-period early-type binary systems might be correlated with the cluster density. However, with its rather large cluster density and low fraction of short-period massive binaries, 
Westerlund\,2 apparently does not follow such a correlation. A more extensive multi-epoch monitoring of the other O-type stars in Westerlund\,2 would be highly valuable in helping us to establish the true binary fraction and investigate the question of a positional dependence of the occurrence of early-type binaries inside the cluster.\\ 

The vast majority of the O-type stars in our sample indicate a RV of about 20 -- 30\,km\,s$^{-1}$ as measured using the He\,{\sc ii} $\lambda$\,5412 line. There are two exceptions to this rule, MSP\,18 and MSP\,171, which are instead found to have negative RVs for this line of ($-1.1 \pm 1.8$)\,km\,s$^{-1}$ and ($-9.3 \pm 6.9$)\,km\,s$^{-1}$ for MSP\,18 and MSP\,171, respectively. Both stars are located outside the dense core of the cluster, at angular separations of 55 and 23\,\arcsec for MSP\,18 and MSP\,171, respectively. Therefore, these objects might have been ejected from the cluster core by means of dynamical interactions (see e.g.\ Leonard \& Duncan \cite{LD}) or be long-period eccentric binaries observed near their quadrature phase. Alternatively, these stars could have strong stellar winds and their He\,{\sc ii} $\lambda$\,5412 line might then be blue-shifted owing to contamination by wind effects. This latter explanation is however at odds with the absence of other strong wind signatures in their spectra (see the discussion of their spectra in the present work and in Paper I). We therefore conclude that these stars are runaway candidates or possible long-period eccentric binaries observed near a phase of maximum RV excursion.\\ 

Finally, we have obtained the very first spectra of mid- to late-type O stars (MSP\,223, A, B, C, D, and E) in Westerlund\,2. Together with the results of Paper I, the number of known O-type stars in the cluster is now 18. However, given the apparent magnitude of these mid to late-type O stars, only the tip of the iceberg has yet been identified. Our current census of O-type stars in Westerlund\,2 is obviously still far from being complete and requires additional high spatial-resolution spectroscopic investigations. Indeed, from the relationship between the mass of the most massive star and the mass of its parental cluster (Weidner et al.\ \cite{Weidner}), we estimate that if $m_{\rm max} = 85\,M_{\odot}$, Westerlund\,2 likely contains about 3000\,$M_{\odot}$ in the form of stars. Adopting the IMF of Weidner et al.\ (\cite{Weidner}), we then estimate that the number of stars more massive than 20\,$M_{\odot}$ should be around 45. Therefore, more than half of the O-type stars of Westerlund\,2 would still await discovery!

\acknowledgement{GR and YN acknowledge support from the Fonds de Recherche Scientifique (FRS/FNRS), through the XMM/INTEGRAL PRODEX contract as well as by the Communaut\'e Fran\c caise de Belgique - Action de recherche concert\'ee - Acad\'emie Wallonie - Europe. We thank Dr.\ Eric Gosset for useful discussions as well as the editor Dr.\ Ralf Napiwotzki and the anonymous referee for their comments and suggestions that helped us improve the manuscript.}

%\listofobjects
\Online

\begin{appendix}
\normalsize
\section{Tables of individual RV measurements}
The tables below provide the heliocentric RVs measured using various lines of stars that do not exhibit any clear binary signature.
\begin{table*}[h!tb]
\caption{Radial velocities of WR\,20b as measured for the various emission lines in the spectrum.\label{RV20b}}
\begin{center}
\begin{tabular}{c c c c c c c c c c}
\hline
Date        & \multicolumn{9}{c}{Radial velocity} \\
HJD-2454000 & H$\beta$ & N\,{\sc v} $\lambda$\,4945 & N\,{\sc iv} $\lambda$\,5204 & He\,{\sc ii} $\lambda$\,5412 & N\,{\sc v} $\lambda$\,5737 & N\,{\sc iv} $\lambda$\,6383 & He\,{\sc ii} $\lambda$\,6406 & H$\alpha$ & He\,{\sc ii} $\lambda$\,6683 \\
& (km\,s$^{-1}$) & (km\,s$^{-1}$) & (km\,s$^{-1}$) & (km\,s$^{-1}$) & (km\,s$^{-1}$) & (km\,s$^{-1}$) & (km\,s$^{-1}$) & (km\,s$^{-1}$) & (km\,s$^{-1}$) \\
\hline
532.587 & 132.2 & $-22.0$ & 57.8 & 108.0 & $-52.9$ &  21.0 & $-15.8$ & 71.1 & 33.9 \\
532.631 & 125.2 & $-23.2$ & 33.6 & 105.9 & $-34.7$ & $-1.2$& $-27.4$ & 71.5 & 41.2 \\
534.581 & 101.9 & $-49.5$ & 36.6 &  87.2 & $-58.8$ &  28.9 & $-30.2$ & 63.3 &   3.1 \\
534.628 &  91.2 & $-20.9$ & 29.0 &  78.7 & $-48.3$ &   7.4 & $-24.5$ & 68.9 & 11.7 \\
536.624 &  84.3 & $-45.0$ & 49.0 &  95.0 & $-72.9$ &  46.8 & $-15.3$ & 11.9 & 44.8\\
537.626 & 125.7 & $-39.4$ & 55.0 & 105.1 & $-38.5$ &$-14.2$&    0.6  & 43.5 & 46.0 \\
538.719 & 132.4 & $-42.2$ & 74.3 & 123.9 & $-58.7$ &  13.5 &    7.4  &100.5 & 66.6 \\
539.622 & 132.1 & $-36.4$ & 64.8 & 125.7 & $-38.9$ &  46.4 &   12.8  &125.8 & 65.0\\
\hline
\end{tabular}
\end{center}
\end{table*}
\begin{table*}[htb]
\caption{Radial velocities of MSP\,18 measured using various absorption lines and the weak C\,{\sc iii} $\lambda$\,5696 emission line.\label{RVMSP18}}
\begin{center}
\begin{tabular}{c c c c c c}
\hline
Date        & \multicolumn{5}{c}{Radial velocity} \\
HJD-2454000 & H$\beta$ & He\,{\sc ii} $\lambda$\,5412 & C\,{\sc iii} $\lambda$\,5696 & H$\alpha$ & He\,{\sc ii} $\lambda$\,6683 \\
& (km\,s$^{-1}$) & (km\,s$^{-1}$) & (km\,s$^{-1}$) & (km\,s$^{-1}$) & (km\,s$^{-1}$)\\
\hline
532.587 & $-41.7$ & $-3.9$ & $-16.8$ & $-122.1$ & 4.2 \\
532.631 & $-46.1$ &   1.7  &  $-0.9$ & $-123.8$ & 9.1 \\
534.581 & $-45.5$ & $-2.1$ & $-17.0$ & $-122.3$ & 9.9 \\
534.628 & $-33.5$ & $-2.0$ & $-17.3$ & $-118.4$ & 14.9 \\
536.624 & $-42.8$ & $-2.0$ &    1.0  & $-122.7$ & 6.3 \\
537.626 & $-34.3$ &   0.6  &  $-8.3$ & $-120.3$ & 19.2 \\
538.719 & $-50.1$ & $-0.5$ &  $-6.8$ & $-122.3$ & 6.4 \\
539.622 & $-44.6$ & $-0.3$ &    7.2  & $-122.5$ & 0.8 \\
\hline
\end{tabular}
\end{center}
\end{table*}
\begin{table*}[htb]
\caption{Radial velocities of MSP\,171 measured using various absorption lines.\label{RVMSP171}}
\begin{center}
\begin{tabular}{c c c c c }
\hline
Date        & \multicolumn{4}{c}{Radial velocity} \\
HJD-2454000 & He\,{\sc ii} $\lambda$\,5412 & O\,{\sc iii} $\lambda$\,5592 & C\,{\sc iv} $\lambda$\,5801 & C\,{\sc iv} $\lambda$\,5812 \\
& (km\,s$^{-1}$) & (km\,s$^{-1}$) & (km\,s$^{-1}$) & (km\,s$^{-1}$)\\
\hline
532.587 &  $-0.1$ & $-10.5$ &   10.3 & $-13.5$ \\
534.581 & $-14.4$ & $-16.9$ & $-7.8$ & $-22.1$ \\
536.624 & $-17.5$ &  $-2.6$ & $-5.0$ & $-17.4$ \\
538.719 &  $-8.6$ & $-23.1$ &   5.5  & $-21.7$ \\
539.622 &  $-5.8$ & $-20.5$ &   5.5  & $-18.3$ \\
\hline
\end{tabular}
\end{center}
\end{table*}
\begin{table*}[htb]
\caption{Same as Table\,\ref{RVMSP171} but for MSP\,182.\label{RVMSP182}}
\begin{center}
\begin{tabular}{c c c c c}
\hline
Date        & \multicolumn{4}{c}{Radial velocity} \\
HJD-2454000 & He\,{\sc ii} $\lambda$\,5412 & O\,{\sc iii} $\lambda$\,5592 & C\,{\sc iv} $\lambda$\,5801 & C\,{\sc iv} $\lambda$\,5812 \\
& (km\,s$^{-1}$) & (km\,s$^{-1}$) & (km\,s$^{-1}$) & (km\,s$^{-1}$) \\
\hline
532.587 & 34.5 & 28.9 & 36.8 &   4.1 \\
534.581 & 27.7 &  --  & 36.5 &   8.0 \\
536.624 & 33.3 & 30.4 & 43.2 &   1.3 \\
538.719 & 29.8 & 29.8 & 38.0 &  16.7 \\
539.622 & 27.0 & 35.6 & 27.5 & $-2.8$ \\
\hline
\end{tabular}
\end{center}
\end{table*}
\begin{table*}[htb]
\caption{Same as Table\,\ref{RVMSP171} but for MSP\,203 + MSP\,444.\label{RVMSP203}}
\begin{center}
\begin{tabular}{c c c c c}
\hline
Date        & \multicolumn{4}{c}{Radial velocity} \\
HJD-2454000 & He\,{\sc ii} $\lambda$\,5412 & O\,{\sc iii} $\lambda$\,5592 & C\,{\sc iv} $\lambda$\,5801 & C\,{\sc iv} $\lambda$\,5812 \\
& (km\,s$^{-1}$) & (km\,s$^{-1}$) & (km\,s$^{-1}$) & (km\,s$^{-1}$) \\
\hline
532.587 & 28.4 & 47.7 & 27.3 & $-0.8$\\
534.581 & 24.7 & 38.8 & 25.6 &   5.5 \\
536.624 & 23.3 & 33.7 & 21.1 &   7.4 \\
538.719 & 21.5 & 39.4 & 17.4 &   0.0 \\
539.622 & 23.9 & 52.6 & 19.8 &   0.6 \\
\hline
\end{tabular}
\end{center}
\end{table*}
\begin{table*}[htb]
\caption{Radial velocities of MSP\,183 measured using various absorption and emission (E) lines.\label{RVMSP183}}
\begin{center}
\begin{tabular}{c c c c }
\hline
Date        & \multicolumn{3}{c}{Radial velocity} \\
HJD-2454000 & He\,{\sc ii} $\lambda$\,5412 & C\,{\sc iv} $\lambda$\,5801 (E) & C\,{\sc iv} $\lambda$\,5812 (E) \\
& (km\,s$^{-1}$) & (km\,s$^{-1}$) & (km\,s$^{-1}$) \\
\hline
532.631 &  23.0 & 38.0 & 53.9 \\
534.628 &  25.3 & 31.3 & 55.5 \\
537.626 &  23.5 & 23.4 & 49.1 \\
\hline
\end{tabular}
\end{center}
\end{table*}
\begin{table*}[htb]
\caption{Radial velocities of MSP\,199, Star B and C measured using the He\,{\sc ii} $\lambda$\,5412 line.\label{RVMSP199}}
\begin{center}
\begin{tabular}{c c c c}
\hline
Date        & \multicolumn{3}{c}{Radial velocity} \\
HJD-2454000 & MSP\,199 & Star B & Star C\\
& (km\,s$^{-1}$) & (km\,s$^{-1}$) & (km\,s$^{-1}$) \\
\hline
532.631 & 29.1 & 18.0 & 28.0 \\
534.628 & 36.4 & 13.6 & 29.7 \\
537.626 & 26.2 & 21.3 & 32.3 \\
\hline
\end{tabular}
\end{center}
\end{table*}
\begin{table*}[htb]
\caption{Same as Table\,\ref{RVMSP171} but for Object A.\label{RVobjA}}
\begin{center}
\begin{tabular}{c c c c c c}
\hline
Date        & \multicolumn{5}{c}{Radial velocity} \\
HJD-2454000 & He\,{\sc i} $\lambda$\,5016 & He\,{\sc ii} $\lambda$\,5412 & O\,{\sc iii} $\lambda$\,5592 & C\,{\sc iv} $\lambda$\,5801 & C\,{\sc iv} $\lambda$\,5812 \\
& (km\,s$^{-1}$) & (km\,s$^{-1}$) & (km\,s$^{-1}$) & (km\,s$^{-1}$) & (km\,s$^{-1}$)\\
\hline
532.631 & 20.6 & 26.3 & 17.1 & 23.0 &  25.6 \\
534.628 &  8.8 & 28.1 & 22.5 & 22.0 &  26.6 \\
537.626 & 19.4 & 28.5 & 16.5 & 18.7 &  19.2 \\
\hline
\end{tabular}
\end{center}
\end{table*}
\end{appendix}
\end{document}